\documentclass[orivec,runningheads]{llncs}

\usepackage{caption}
\usepackage{subcaption}

\usepackage{graphicx}

\usepackage[table]{xcolor}

\usepackage{cite}

\usepackage[noend,linesnumbered,ruled]{algorithm2e}

\usepackage{xspace}

\usepackage{multirow}
\usepackage{booktabs}

\usepackage{amsmath, amsfonts, amssymb}
\usepackage{cancel}
\usepackage{mathtools}

\usepackage{thmtools}
\usepackage{thm-restate}

\usepackage{wasysym}

\usepackage{wrapfig}

\usepackage{orcidlink}

\usepackage[inline]{enumitem}

\usepackage[skins]{tcolorbox}

\usepackage{tikz}

\usepackage{hyperref}

\usepackage[capitalise,english,nameinlink]{cleveref} 
\crefname{line}{\text{line}}{\text{lines}} %

\definecolor{cispa-blue}{HTML}{009CC4}
\definecolor{cispa-blue-80}{HTML}{33B0D0}
\definecolor{cispa-blue-60}{HTML}{66C4DC}
\definecolor{cispa-blue-40}{HTML}{99D7E7}
\definecolor{cispa-blue-20}{HTML}{D5ECF6}
\definecolor{cispa-dark-blue}{HTML}{002864}
\definecolor{cispa-medium-blue}{HTML}{018DCA}
\definecolor{cispa-bright-blue}{HTML}{00C8FF}
\definecolor{cispa-anthracite}{HTML}{1C1C1C}
\definecolor{cispa-dark-gray}{HTML}{2B2B2B}
\definecolor{cispa-light-gray}{HTML}{C9C9C9}
\definecolor{cispa-grid-gray-light}{HTML}{E8E8E8}
\definecolor{cispa-offwhite}{HTML}{F5F5F5}
\definecolor{cispa-green}{HTML}{008040}
\definecolor{cispa-red}{HTML}{EB3300}
\definecolor{cispa-orange}{HTML}{FF7F32}
\definecolor{cispa-yellow}{HTML}{FFC600}
\definecolor{cispa-gray}{HTML}{969595}
\definecolor{cispa-blue-10}{HTML}{EBF5F9}
\definecolor{cispa-blue-gray-dark}{HTML}{004C61}
\definecolor{cispa-blue-gray-light}{HTML}{6D8995}
\definecolor{cispa-blue-green}{HTML}{00C49A}
\definecolor{cispa-blue-green-20}{HTML}{DBF4EA}
\definecolor{cispa-blue-green-bright}{HTML}{00EEBB}

\usetikzlibrary{automata, arrows.meta, positioning}

\tikzstyle{lts}=[auto, ->, >=stealth, font=\small]

\tikzstyle{every node}=[initial text=, font=\small]

\tikzstyle{location}=[
    rectangle,
    rounded corners,
    minimum size=14pt,
    draw=cispa-anthracite,
    fill=cispa-blue-20,
    text=cispa-anthracite,
    inner sep=3.5pt
]

\tikzstyle{other-location}=[
    rectangle,
    rounded corners,
    minimum size=14pt,
    draw=cispa-anthracite,
    fill=cispa-grid-gray-light,
    text=cispa-anthracite,
    inner sep=3.5pt
]

\tikzstyle{invariant}=[
    fill=cispa-yellow,
    inner sep=1pt,
    node distance=1.5pt,
    font=\itshape\scriptsize,
]

\tikzstyle{locguard}=[
    fill=cispa-blue,
    inner sep=0.5pt,
    node distance=2pt,
    yshift=-0.5pt,
    font=\scriptsize,
]

\tikzstyle{final}=[double]

\tikzstyle{init}=[initial, initial distance=0.35cm]

\ifx\homepage\undefined
\usepackage{fontawesome}
\newcommand{\homepage}[1]{\href{#1}{\color{gray}\faHome}}
\fi

\hypersetup{
    colorlinks = true,
    breaklinks = true,
    citecolor   = cispa-dark-blue,
    linkcolor   = cispa-dark-blue,
    urlcolor    = cispa-medium-blue,
}

\newcommand{\eg}{e.g.,\xspace}

\newcommand{\ie}{i.e.,\xspace}

\newcommand{\smartpar}[1]{\noindent \textbf{#1}.}

\usepackage{pifont} %
\newcommand{\xmark}{\leavevmode{\color{red}\ding{55}}}%
\newcommand{\cmark}{\leavevmode{\color{green!75!black}\ding{51}}}%

\newcommand{\presbCite}{\cite{Scarpellini84,BoroshT76,GathenS78,Haase18}}

\newcommand{\NP}{\ensuremath{\textsc{NP}}\xspace}

\newcommand{\coNEXP}{\ensuremath{\textsc{coNEXP}}\xspace}
\newcommand{\NEXP}{\ensuremath{\textsc{NEXP}}\xspace}

\newcommand{\Nat}{\ensuremath{\mathbb{N}}}
\newcommand{\unitvec}[1]{\ensuremath{\textbf{u}_{#1}}}

\newcommand{\initial}[1]{\ensuremath{#1_{\mathcal{I}}}}
\renewcommand{\max}[1]{\ensuremath{\mathsf{max}(#1)}}
\renewcommand{\min}[1]{\ensuremath{\mathsf{min}(#1)}}

\newcommand{\globally}{\ensuremath{\square}}

\newcommand{\localActs}{\ensuremath{\Sigma}}
\newcommand{\globalActs}{\ensuremath{\mathbf{\Sigma}}}

\newcommand{\actdec}[1]{\ensuremath{\mathtt{#1}}}
\newcommand{\act}{\ensuremath{\actdec{\actdec{a}}}}

\newcommand{\run}{\ensuremath{\rho}}

\newcommand{\Locs}{\ensuremath{\mathcal{L}}}
\newcommand{\loc}{\ensuremath{l}}

\newcommand{\trans}[1]{\ensuremath{\xrightarrow{#1}}}

\newcommand{\config}{\ensuremath{\textbf{c}}}
\newcommand{\Configs}{\ensuremath{\textbf{C}}}

\newcommand{\CS}[1]{\ensuremath{\mathsf{CS}(#1)}}

\newcommand{\ta}{\ensuremath{A}}
\newcommand{\Params}{\ensuremath{\Pi}}
\newcommand{\param}{\ensuremath{p}}
\newcommand{\SharedVars}{\ensuremath{\Gamma}}
\newcommand{\Rules}{\ensuremath{\mathcal{R}}}
\newcommand{\RC}{\ensuremath{RC}}
\newcommand{\coef}{\ensuremath{q}}

\newcommand{\rl}{\ensuremath{r}}
\newcommand{\vfrom}{\ensuremath{\text{from}}}
\newcommand{\vto}{\ensuremath{\text{to}}}

\newcommand{\vup}{\ensuremath{\mathbf{uv}}}
\newcommand{\vk}{\ensuremath{\mathbf{k}}}
\newcommand{\vg}{\ensuremath{\mathbf{g}}}
\newcommand{\vp}{\ensuremath{\mathbf{p}}}
\newcommand{\guard}{\ensuremath{\varphi}}

\newcommand{\N}{\ensuremath{\mathcal{N}}}
\newcommand{\Np}{\ensuremath{\mathcal{N}\!(\mathbf{p})}}

\newcommand{\env}{\ensuremath{\mathcal{E}}}

\newcommand{\var}{\ensuremath{x}}
\newcommand{\varinc}{\ensuremath{\text{\small\texttt{++}}}}

\newcommand{\TAS}{TAS\xspace}
\newcommand{\globalRules}{\ensuremath{\mathcal{G}}}

\newcommand{\hi}{\ensuremath{\text{high}}}
\newcommand{\low}{\ensuremath{\text{low}}}

\newcommand{\sink}{\ensuremath{\oslash}}
\newcommand{\sync}{\ensuremath{\mathbf{s}}}
\newcommand{\accfac}{\ensuremath{n}}
\newcommand{\Accfacs}{\ensuremath{N}}
\newcommand{\localfac}{\ensuremath{r}}

\newcommand{\syncta}{\ensuremath{\mathcal{A}}}
\newcommand{\noact}{\ensuremath{\epsilon}}

\newcommand{\tsender}{\ensuremath{\mathsf{each}}\xspace}
\newcommand{\tmaximal}{\ensuremath{\mathsf{max}}\xspace}

\newcommand{\send}{\ensuremath{\text{Send}}}
\newcommand{\recv}{\ensuremath{\text{Recv}}}
\newcommand{\type}{\ensuremath{\mathsf{type}}\xspace}

\newcommand{\layer}{\ensuremath{L}}
\newcommand{\sinkLayer}{\ensuremath{\layer_\sink}}
\newcommand{\layeridx}{\ensuremath{\ell}}

\newcommand{\someact}{\ensuremath{\alpha}}

\newcommand{\player}{\ensuremath{P}}

\newcommand{\cardConstraint}{\ensuremath{\varphi}}

\newcommand{\phisuc}{\ensuremath{\phi_{\mathsf{suc}}}}
\newcommand{\philoc}{\ensuremath{\phi_{\mathsf{loc}}}}
\newcommand{\phiacc}{\ensuremath{\phi_{\mathsf{acc}}}}

\newcommand{\phisend}{\ensuremath{\phi_{\mathsf{each}}}}
\newcommand{\phimax}{\ensuremath{\phi_{\mathsf{max}}}}
\newcommand{\phisync}{\ensuremath{\phi_{\mathsf{sync}}}}
\newcommand{\phireach}{\ensuremath{\phi_{\mathsf{reach}}}}
\newcommand{\phireachta}{\ensuremath{\phi_{\mathsf{asyn}}}}

\newcommand{\phieq}{\ensuremath{\phi_{\mathsf{rp}}}}
\newcommand{\phieqstrong}{\ensuremath{\phi_{\mathsf{srp}}}}
\newcommand{\phiinit}{\ensuremath{\initial{\phi}}}

\newcommand{\Part}{\ensuremath{\mathbb{P}}}
\newcommand{\concat}{\ensuremath{::}}

\newcommand{\LocsRM}{\ensuremath{\Locs_{rm}}}
\newcommand{\VarsRM}{\ensuremath{\SharedVars_{rm}}}
\newcommand{\LocsAdd}{\ensuremath{\Locs_{add}}}

\newcommand{\LocsIn}{\ensuremath{\Locs_{ent}}}
\newcommand{\LocsOut}{\ensuremath{\Locs_{ext}}}
\newcommand{\RulesAdd}{\ensuremath{\Rules_{add}}}
\newcommand{\globalRulesAdd}{\ensuremath{\mathcal{G}_{add}}}
\newcommand{\replacement}{\ensuremath{\mathbf{R}}}

\newcommand{\replace}[2]{\ensuremath{#1 \lhd #2}}

\newcommand{\replaceTA}[2]{\ensuremath{\syncta_{\mathsf{chng}}({#1},{#2})} }

\begin{document}
\title{Modular Parameterized Model Checking\\ of Threshold-Based Distributed Algorithms}
\titlerunning{Modular Parameterized Model Checking}

\author{
    Paul Eichler\inst{1}\orcidlink{0009-0008-6117-318X}\and
    Swen Jacobs\inst{1}\orcidlink{0000-0002-9051-4050}
}
\authorrunning{P. Eichler \and S. Jacobs}

\institute{
    CISPA Helmholtz Center for Information Security, Saarbr\"ucken, Germany\\
    \email{\{ paul.eichler, jacobs \}@cispa.de}
}

\maketitle

\begin{abstract}
    We present a novel approach for model checking fault-tolerant distributed algorithms:
    instead of building large, monolithic algorithm models, we propose a modular approach where the user decomposes the algorithm into modules, verifies them individually, and replaces them with a synchronous abstraction.
    To enable our technique, we introduce a new system model that combines aspects of threshold automata and global synchronization protocols.
    We investigate the expressiveness of our new model, and present a (bounded) model checking procedure as well as a method for checking the correctness of the abstraction step.
    We implemented the technique and report on promising experimental results on a case study.

\end{abstract}

\section{Introduction} %

Large-scale distributed systems, such as cloud platforms or blockchain systems, increasingly form the backbone of much of our industry and society. 
Bugs or failures in these systems can lead to significant harm, which encouraged a wide range of efforts to formally verify their correctness~\cite{HawblitzelHKLPR17,JaberWJKS21,McMillanP20,RahliGBC15,WilcoxWPTWEA15,PirleaGKZS25,OtoniFKKMOPTK26}.

However, the task remains challenging because of the intrinsic complexity of the setting.
Distributed systems and their underlying algorithms are usually required to scale to an arbitrary number of participants.
Since the probability of a failure in such a system increases with the number of components, to be practical, distributed algorithms also need to be fault-tolerant.
In settings like blockchain protocols, where arbitrary untrusted nodes can join, the algorithms need to withstand arbitrary failures referred to as \emph{Byzantine failures}~\cite{LamportSP82}. 

One of the main techniques for verifying the correctness of distributed algorithms is \emph{parameterized model checking}~\cite{Esparza14,Bloem2015,AbdullaST18}, where the number of participants is treated as a parameter and verification is conducted for any parameter value.
Even though the parameterized model checking problem (PMCP) is undecidable in general \cite{AptK86}, even for very restricted classes of systems~\cite{Suzuki88}, many system classes for which the problem is decidable have been identified~\cite{GermanS92,EsparzaFM99,EmersonK00,EmersonK03}.
In particular, threshold automata (TA)~\cite{KonnovLVW17,KonnovVW17,fmsdKonnovLVW17} provide a framework for the modeling and verification of (threshold-based) distributed algorithms with Byzantine fault-tolerance, where processes communicate via shared variables to model message-passing, and decisions can depend on these shared variables meeting certain (parametric) threshold values.
TA have been used to verify correctness of a wide range of consensus and broadcast algorithms~\cite{BrasileiroGMR01,MostefaouiMPR03,BrachaT85,SPIN13,BertrandKLW19,CrainNG21,0001LSW23}.

A major challenge of parameterized model checking is the requirement to create a faithful model of the algorithm under consideration.
Since distributed algorithms can be very complex, the resulting models can become so large that they are effectively impossible to understand or mechanically verify, even in a decidable fragment.
A potential solution to this problem is to model such systems at different levels of granularity in order to keep the verification problem manageable at every separate level. 

Global synchronization protocols (GSPs)~\cite{JaberJW0S20,JaberWJKS21,JaberWJKS23} are a suitable high-level model for (applications built on) distributed algorithms, abstracting from individual message passing between processes by summarizing the result as a synchronous global step.\footnote{Note that this is different from synchronous threshold automata~\cite{StoilkovskaKWZ19,StoilkovskaKWZ21,StoilkovskaKWZ22}, where processes execute in lock-step, but the algorithm is still modeled on the level of individual message passing.} 
However, modeling an application as a GSP requires hard trade-offs in precision.
In particular, the effects of Byzantine failures are not modeled explicitly, but are \emph{assumed} to be completely encapsulated in synchronous steps.
Moreover, modeling on different levels of precision requires additional correctness arguments connecting the different levels, which may be hard to obtain.

To facilitate truly modular parameterized model checking, this work proposes a system model that supports both fine-grained modeling of individual process steps and high-level synchronous steps, with explicit support of Byzantine faults.
This enables the modeling of complex applications at different levels of abstraction within the same framework, using \emph{synchronous module abstraction} to replace multiple asynchronous steps by one synchronous step.
We determine conditions under which parameterized model checking problems in our setting are decidable, and show how to automatically verify the correctness of a synchronous module abstraction. 
To the best of our knowledge, this is the first \emph{modular} parameterized model checking approach that directly supports different levels of abstraction, and automatically produces a correctness argument for the abstraction.

\smartpar{Contributions} 
In this paper, we introduce the model of \emph{threshold automata with synchronous actions}, which is an extension of TA with global steps that are inspired by GSPs, but adapted to a setting with shared variables and Byzantine fault-tolerance (\cref{sec:tas}).
We show that adding this type of synchronous actions to TA in general makes the PMCP undecidable, even for simple specifications like state reachability (\cref{sec:complexity}).
In \cref{sec:reocver-dec}, we establish a bounded model checking procedure, which becomes a decision procedure if we can statically bound the number of synchronous steps in any execution of the protocol. We argue that this will be the case for many of our intended applications. 

We then introduce the technique of \emph{synchronous module abstraction}, which combines our bounded model checking technique at different levels of abstraction with an automatic correctness check for the abstractions (\cref{sec:modular-replace}), and demonstrate how to apply it to a case study. 
Finally, we report on a prototype implementation of our technique and promising experimental results on the modeling and verification of a modular king consensus algorithm (\cref{sec:eval}).

\section{Preliminaries}
\label{sec:prelims}
Before presenting our new system model, we recapitulate the main definitions for threshold automata, an existing model for Byzantine fault-tolerant algorithms, \ie algorithms where faulty nodes can behave maliciously.

Threshold automata communicate via shared integer variables. 
Their transitions can be guarded with constraints over these variables and the parameters.
Parameters are determined by the environment and used to, for example, represent the number of participating processes (usually $n$) or the bound on the number of faulty processes (usually $t$).

\begin{definition}[Environment~\cite{Balasubramanian20}]
    An \emph{environment} of a TA is a tuple $\env = \langle\Params, \RC, \N\rangle$, where $\Params$ denotes the set of \emph{parameters}, which are variables over $\Nat_0$, $\RC$ is the \emph{resilience condition}, a linear arithmetic formula over the  parameters, and $\N:\Params_{\RC} \rightarrow \Nat_0$ a linear function, where $\Params_{\RC}$ are the parameter valuations that satisfy $\RC$.
\end{definition}

The resilience condition reflects assumptions on the parameters, for example bounding the number of faults.
For the Byzantine fault model, a typical resilience condition is $n > 3t$. 
The function $\N$ determines the number of correct processes. 
If not stated otherwise we assume $\Np = n {-} f$.

\begin{definition}[Threshold Automaton~\cite{KonnovVW14,KonnovVW17}]\label{def:ta}
    A \emph{threshold automaton} over environment $\env = \langle\Params, \RC, \N\rangle$ is a tuple $\ta = \langle \Locs, \initial{\Locs}, \SharedVars, \Rules \rangle$ where
    $\Locs$ is a finite set of locations,
    with $\initial{\Locs} \subseteq \Locs$ the initial locations,
    $\SharedVars$ the set of shared variables over $\Nat_0$, and $\Rules$ is a set of rules.
    A \emph{rule} is a tuple $\rl = (\vfrom, \vto, \varphi, \vup)$, where  $\vfrom,\vto \in \Locs$, $\vup \in \Nat_0^{|\SharedVars|}$ the  \emph{update vector} for shared variables and $\varphi$ is a conjunction of lower guards and upper guards.
    A \emph{lower guard} is a constraint $\coef_0 {+} \sum_{i=1}^{|\Params|} \coef_i {\cdot} \param_i \leq x$, with $x \in \SharedVars$, $\coef_0,\ldots,\coef_{|\Params|} \in \mathbb{Z}$, $\param_1,\ldots, \param_{|\Params|} \in \Params$.
    An \emph{upper guard} is a constraint $\coef_0 {+} \sum_{i=1}^{|\Params|} \coef_i {\cdot} p_i > x$.
    The left-hand side of a guard is called a \emph{threshold}.
\end{definition}

\begin{example}\label{ex:weak-cons-ta-intro}
    \begin{figure}
        \hfill
        \begin{minipage}[c][32mm]{0.48\textwidth}
            \centering
            \resizebox{\textwidth}{!}{
                \input{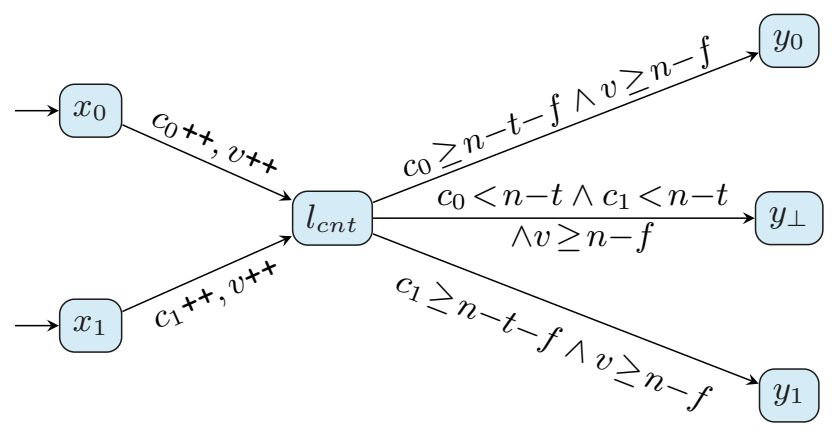}
            }
        \end{minipage}
        \hfill
        \begin{minipage}[c][32mm]{0.48\textwidth}
            \centering
            \resizebox{0.98\textwidth}{!}{
                \begin{tikzpicture}[lts]
    \node (i0) [location,initial, initial text=] at (0, 1) {$x_0$};
    \node (i1) [location,initial, initial text=] at (0, -1) {$x_1$};

    \node (cnt) [location] at (2.25,0) {$\loc_{cnt}$};

    \node (o0) [location] at (6.5, 1.65) {$y_0$};
    \node (of) [location] at (6.5, 0) {$y_\bot$};
    \node (o1) [location] at (6.5, -1.65) {$y_{1}$};

    \path[->]
        (i0) edge [] node [above, sloped,yshift=-2pt] {$c_0\varinc,v\varinc$} (cnt)
        (i1) edge [] node [below, sloped] {$c_1\varinc,v\varinc$} (cnt)

        (cnt) edge [] node [above, sloped,yshift=-2pt] {$c_0 \!\geq\! n{-}t{-}f \wedge v \!\geq\! n{-}f$} (o0)
        (cnt) edge [] node [below, sloped,yshift=-2pt] {$c_1 \!\geq\! n{-}t{-}f \wedge v \!\geq\! n{-}f$} (o1)

        (cnt) edge [] node [above, sloped,xshift=5pt,yshift=-12.5pt,align=center] {$c_0 \!<\! n{-}t \wedge c_1 \!<\! n{-}t$\\$\wedge v \!\geq\! n{-}f$} (of)
    ;
\end{tikzpicture}
                \vspace{10mm}
            }
            \caption{Weak consensus threshold automaton $A_{WC}$, with $\Params = \{ n, t, f \}$, and $\RC = n > 3t \wedge t \geq f$}
            \label{fig:weak-cons-ta}
        \end{minipage}
        \hfill
    \end{figure}

    \cref{alg:weak-cons} shows the weak consensus algorithm from \cite{FitziM00,Dolev82}.
    Each process $P_i$ starts with initial value $x_i \in \{0, 1\}$ and the goal is to either agree on value $0$ or $1$ or return $\bot$ if no agreement is reached.
    To this end, each process sends its input value $x_i$ to all other processes and counts how many votes it received for each value.
    If any value gets at least $n{-}t$ votes, it will be written to $y_i$, otherwise $y_i$ is set to $\bot$.

    The threshold automaton $A_{WC}$ modelling the algorithm is presented in \cref{fig:weak-cons-ta}.
    The two shared variables $c_0, c_1$ count the votes for each value, and variable $v$ tracks the overall number of votes.
    The parameter $n$ represents the number of processes, $t$ bounds the number of faults, and $f$ is the actual number of faults.
    All (correct) processes start in $x_{0}$ or $x_{1}$ depending on their initial value.
    They then move into $\loc_{cnt}$ incrementing $v$ and either $c_0$ or $c_1$, and can transition to $y_{0}$ or $y_{1}$ if the corresponding variable reaches the threshold $n{-}t{-}f$ and $n{-}f$ votes have been cast.
    Otherwise, they can only move on to location $y_\bot$.

    Note that the TA here models only the honest processes, and parameter $f$ is used in thresholds to model the effects of Byzantine faults. 
    In particular, since Byzantine processes can send votes for any value, an honest process may see up to $f$ additional votes for any value, reflected by the threshold $n{-}t{-}f$.
\end{example}

\smartpar{Semantics of TA}
The TA model assumes that $\Np$ honest processes are running concurrently, and that the effects of faulty processes are modeled in the guard conditions. 
Let $\unitvec{n}$ denote the unit vector, which is one at position $n$ and zero otherwise.
The semantics of TA is then defined by their counter system representation:

\begin{definition}[TA Counter System \cite{EmersonT99,KonnovVW14,KonnovVW17,BaumeisterEJSV24}]\label{def:ta-cs}
    The counter system of a TA $\ta = \langle \Locs, \initial{\Locs}, \SharedVars, \Rules \rangle$ under environment $\env = \langle \Params, \RC, \N \rangle$ is the tuple $\CS{\env,\ta} =\langle \Configs, \initial{\Configs}, \trans{} \rangle$ where
    \begin{itemize}
        \item $\Configs$ is the set of configurations. A \emph{configuration} is a triple $\config = (\vk, \vg, \vp)$ where
              \begin{itemize}
                  \item $\vk \in \Nat_0^{|\Locs|}$ assigns a number of processes to each location.
                        We use $\config.\vk[\loc]$ to denote the processes currently in location $\loc$.
                  \item $\vg \in \Nat_0^{|\SharedVars|}$ is the valuation of the shared variables.
                        For $x \in \SharedVars$, we use $\config.\vg[x]$ to denote the valuation of $x$ in $\config$.
                  \item $\vp \in \Nat_0^{|\Params|}$ is a parameter valuation such that $\vp \models \RC$, \ie $\RC$ holds after substituting each parameter with its value in $\vp$.
              \end{itemize}
        \item The set of initial configurations $\initial{\Configs}$ contains all initial configurations, \ie configurations $\config$ that satisfy
              \[
                  \forall x \in \SharedVars: \config.\vg[x] = 0  \wedge \sum\limits_{\loc \in \initial{\Locs}}\!\!\config.\vk[\loc] = \N(\config.\vp) \wedge  \forall \loc \notin \initial{\Locs}.\config.\vk[\loc] = 0
              \]
        \item The transition relation $\trans{} \subseteq \Configs \times \Rules \times \Configs$ contains a transition $(\config,\rl,\config')$ (denoted as $\config \trans{\rl} \config'$) iff the following conditions hold:
              \begin{itemize}
                  \item $\config.\vp = \config'.\vp$, \ie the valuation of parameters did not change.
                  \item $\config.\vk[\rl.\vfrom] > 0$ and $\config'.\vk = \config.\vk - \unitvec{\rl.\vfrom} + \unitvec{\rl.\vto}$, \ie a process moves from $\rl.\vfrom$ to $\rl.\vto$.
                  \item $\config.\vg \models \rl.\guard$, \ie $\guard$ holds after replacing shared variables with values $\config.\vg$
                  \item $\config'.\vg = \config.\vg + \rl.\vup$
              \end{itemize}
    \end{itemize}
\end{definition}

\smartpar{Run \& Computation}
The execution of a transition $\config \trans{\rl} \config'$ is called a \emph{step} from $\config$ to $\config'$ using rule $\rl$.
A \emph{run $\run$} of $\CS{\env,\ta}$ is a sequence of configurations and rules $\config_0 \rl_0 \config_1 \dots \rl_{n-1} \config_n$, where for every $i \in 0,...,n - 1$ there exists a rule $\rl_i \in \Rules$, such that $\config_i \trans{\rl} \config_{i+1}$.
We call $\run$ a \emph{computation} if $\config_0 \in \initial{\Configs}$.
A configuration $\config$ is \emph{reachable} if there exists a computation that ends in $\config$.

\section{Threshold Automata with Synchronous Actions}\label{sec:tas}

To motivate our new model with synchronous actions, we start with an example that highlights challenges when modeling a modular algorithm as a TA.

\begin{example}\label{ex:graded-cons}
    \begin{figure}
        \hfill
        \begin{minipage}[c][43mm]{0.49\textwidth}
            \centering
            \resizebox{\textwidth}{!}{
                \input{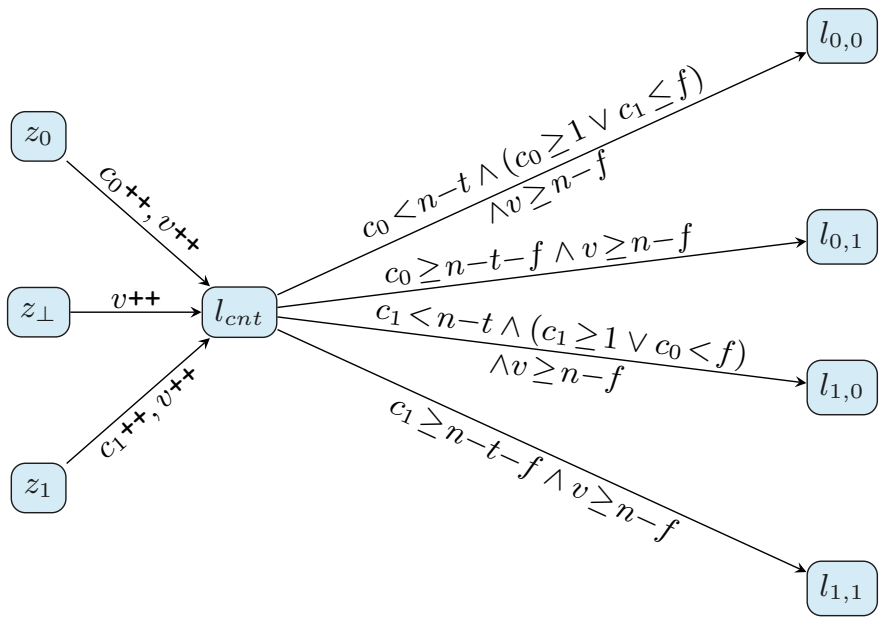}
            }
        \end{minipage}
        \hfill
        \begin{minipage}[c][43mm]{0.49\textwidth}
            \centering
            \resizebox{\textwidth}{!}{
                \begin{tikzpicture}[lts]
    \node (i0) [location] at (1, 1.75) {$z_{0}$};
    \node (if) [location] at (1, 0) {$z_{\bot}$};
    \node (i1) [location] at (1, -1.75) {$z_{1}$};

    \node (cnt) [location] at (3,0) {$\loc_{cnt}$};

    \node (o00) [location] at (9, 2.75) {$\loc_{0,0}$};
    \node (o01) [location] at (9, 0.75) {$\loc_{0,1}$};
    \node (o10) [location] at (9, -0.75) {$\loc_{1,0}$};
    \node (o11) [location] at (9,-2.75) {$\loc_{1,1}$};

    \path[->]
        (i0) edge [] node [above, sloped,yshift=-2pt] {$c_0\varinc,v\varinc$} (cnt)
        (i1) edge [] node [below, sloped,yshift=1.5pt] {$c_1\varinc,v\varinc$} (cnt)
        (if) edge [] node [above, sloped,yshift=-2pt] {$v\varinc$} (cnt)
    ;

    \path[->]
        (cnt) edge [] node [above,sloped,yshift=-12pt, align=center] {$c_0 \!<\!n{-}t \wedge (c_0 \!\geq\! 1 \vee c_1 \!\leq\! f)$\\$\wedge v \!\geq\! n{-}f$} (o00)
        (cnt) edge [] node [above,sloped,yshift=-2pt] {$c_0 \!\geq\! n{-}t{-}f \wedge v \!\geq\! n {-} f$} (o01)
        (cnt) edge [] node [above,sloped,xshift=5pt,yshift=-13pt,align=center] {$c_1 \!<\! n{-}t \wedge (c_1 \!\geq\! 1 \vee c_0 \!<\! f)$\\$\wedge v \!\geq\! n{-}f$} (o10)
        (cnt) edge [] node [below,sloped,yshift=1.5pt] {$c_1 \!\geq\! n{-}t{-}f \wedge v \!\geq\! n {-} f$} (o11)
    ;
\end{tikzpicture}
                \vspace{10mm}
            }
            \caption{Graded-consensus threshold automaton $A_{GC}$, with $\Params = \{ n, t, f \}$ and $\RC = n > 3t \wedge t \geq f$}
            \label{fig:graded-cons-ta}
            \vfill
        \end{minipage}
        \hfill
    \end{figure}

    \cref{alg:graded-cons} shows the pseudocode of graded consensus~\cite{FitziM00} which uses weak consensus (\cref{ex:weak-cons-ta-intro}) as a module.
    The algorithm calls weak consensus, re-shares the computed result among all processes, and decides on a value $0$ or $1$, depending on which value the majority accepted.
    Additionally, it assigns a grade to the result, which is set to $1$ if at least $n {-} t$ processes accepted the chosen value in the weak consensus step, and $0$ otherwise.

    The TA $A_{GC}$ depicted in \cref{fig:graded-cons-ta} models graded consensus except for the call to weak consensus.
    Instead, the initial states $z_{0},z_{1}$ and $z_{\bot}$ correspond to the final locations $y_{0}, y_{1}$ and $y_\bot$ from the TA for weak consensus (\cref{fig:weak-cons-ta}), and a full description of graded consensus is obtained by appending $A_{GC}$ to $A_{WC}$, identifying the respective locations (see \cref{fig:replace-graded-before}).
\end{example}

\cref{ex:graded-cons} shows that, while the pseudo code uses weak consensus as an abstract module, the TA representation does not have an equivalent mechanism, and instead requires including a full copy of weak consensus.
We will see that graded consensus is in turn a module of the king consensus algorithm, the TA for king consensus will include the full TAs of both weak and graded consensus.
This leads to complex models that are difficult to check (for details see \Cref{sec:eval}).

We propose a different approach that allows us to abstract modules into synchronous steps and thus reduce the complexity of the overall model.

\subsection{Formal Model}\label{sec:formal-model}

The synchronous actions we introduce are inspired by broadcast protocols~\cite{EmersonN98,EsparzaFM99} and global synchronization protocols (GSPs)~\cite{JaberJW0S20,JaberWJKS21}.
A threshold automaton with synchronous actions is a TA that, in addition to the regular asynchronous transitions, can have \emph{synchronous actions} where all processes take a step at the same time.
We define a finite set of \emph{global actions} $\globalActs$. 

\begin{definition}[Threshold Automaton with Synchronous Actions]
    \label{def:TAS}
    A \emph{threshold automaton with synchronous actions (\TAS)} is a tuple $\syncta = \langle\Locs, \initial{\Locs}, \SharedVars, \allowbreak\Rules, \globalRules\rangle$, where $\Locs, \initial{\Locs}, \SharedVars, \Rules$ are as in \cref{def:ta} and $\globalRules$ contains for every $\act \in \globalActs$ one 
     \emph{global rule} $(\act,\send,\recv,\type,\low,\hi)$, where
\begin{itemize}
    \item $\send$ is a finite set of \emph{local send transitions} $(\loc_i,\act_i!!,\loc_i')$, with $0 \leq i < k$ for some $k \in \Nat$.
    \item $\recv$ is a finite set of \emph{local receive transitions} of the form $(\loc, \act??, \loc')$. %
    \item \type is either \emph{\tsender} or \emph{\tmaximal}.
          Intuitively, if \type is \tsender, then each sending transition needs to be taken by at least one process, and if it is \tmaximal then the maximum number of available processes need to take the sending transitions.
    \item the \emph{lower bound} $\low$ and the \emph{upper bound} $\hi$ are given as thresholds (as defined in \cref{def:ta}) and define (in addition to \type) the minimum and maximum numbers of senders that can execute $\act$ concurrently.
\end{itemize}
\end{definition}

For a given action $\act \in \globalActs$, we will refer to the components of the global rule as $\act.\send$, $\act.\recv$, etc. and write $\loc \trans{\someact} \loc'$ if there exists a global rule with some action $\act$ such that $(\loc, \act_i!!, \loc') \in \act.\send$ or $(\loc,\act??,\loc') \in \act.\recv$.
Local send transitions and local receive transitions will also be called \emph{synchronous transitions}.
Labels $\act_i!!$ will be called \emph{local send actions}, and $\act??$ \emph{local receive actions}.

To separate them from the global rules, we will refer to rules from $\Rules$ as \emph{local rules} and write $\loc \trans{\noact} \loc'$ if there exists $r \in \Rules$ with $r.\vfrom = \loc$ and $r.\vto = \loc'$.

\addvspace{7pt plus 2pt minus 4pt}%
\begin{wrapfigure}[16]{r}{0.45\linewidth}
\vspace{-18pt}
\scalebox{0.75}{
\begin{tikzpicture}[lts]
    \node (i0) [location,init] at (0, 0.75) {$x_{0}$};
    \node (i1) [location,init] at (0, -0.75) {$x_{1}$};

    \node (o0) [location] at (6,1.25) {$y_{0}$};
    \node (of) [location] at (6, 0) {$y_\bot$};
    \node (o1) [location] at (6,-1.25) {$y_{1}$};

    \path[->]
        (i0) edge [bend left=15] node [above, sloped,yshift=-2pt] {$\actdec{z}!!,\actdec{mz}!!,\actdec{z}??$} (o0)
        (i1) edge [bend right=15] node [below, sloped,yshift=1.5pt] {$\actdec{o}!!,\actdec{mo}!!,\actdec{o}??$} (o1)

        (i1) edge [bend left=12] node [sloped,xshift=15pt,yshift=-2pt] {$\actdec{z}??,\actdec{mz}??$} (o0)
        (i0) edge [bend right=12] node [below,sloped,xshift=15pt,yshift=1.5pt] {$\actdec{o}??,\actdec{mo}??$} (o1) 

        (i0) edge [] node [above, sloped,xshift=30pt,yshift=-2pt] {$\actdec{mz}!!,\actdec{b}!!,\actdec{mo}??$} (of)
        (i1) edge [] node [below, sloped,xshift=30pt,yshift=1.5pt] {$\actdec{mo}!!,\actdec{mz}??,\actdec{b}??$} (of)
    ;
\end{tikzpicture}
}
\caption{\TAS $\syncta_{WC}$ of \cref{alg:weak-cons} with $\actdec{z},\actdec{o},\actdec{b}$ \tsender actions with $\actdec{z}.\low = \actdec{o}.\low = n{-}t{-}f$, $\actdec{b}.\low = 1$, $\actdec{b}.\hi = n{-}t{-}f{-}1$ and $\actdec{mz}, \actdec{mo}$ \tmaximal actions with $\actdec{mo}.\low = \actdec{mz}.\low = n{-}t{-}f$ and $\actdec{mz}.\hi = \actdec{mo}.\hi = n{-}t{-}1$.}
        \label{fig:weak-cons-sync}
\end{wrapfigure}

\noindent{\refstepcounter{example}\label{ex:weak-cons-sync}}\textit{Example~\theexample.}\ %
    \cref{fig:weak-cons-sync} depicts a model of the weak consensus algorithm from \cref{ex:weak-cons-ta-intro} using only synchronous actions.
    Global action $\actdec{z}$ is an \tsender action with local sending action $\actdec{z}!!$ which transitions from $x_{0}$ to $y_{0}$ and local receiving actions $\actdec{z}??$ that transition to $y_{0}$.
    The action $\actdec{z}$ is enabled if at least $n{-}t{-}f$ ($\actdec{z}.\low$) processes start in $x_{0}$ and when $\actdec{z}$ is executed, all processes move to $y_{0}$.
    Action $\actdec{z}$ intuitively represents acceptance of value zero by all processes.
    The action $\actdec{o}$ is the analog action for accepting value one.

    Actions $\actdec{mo}$ and $\actdec{mz}$ represent an execution where no value has $n{-}t$, therefore the upper bound, but enough such that with votes by the faulty processes some process could accept a value.
    The action $\actdec{b}$ represents the case where no value can be accepted (even with votes from faulty processes).
    Therefore, all processes must transition to $y_\bot$.

    Note that, in the depicted automaton, not every location has a receive action for every global action.
    This will be discussed later in more detail.
\par\addvspace{7pt plus 2pt minus 4pt}%

\smartpar{Semantics of TA with Synchronous Actions}\label{sec:semantics}
In a \TAS $\syncta$, the transition relation from \Cref{def:ta-cs} is extended with global transitions $(\config,\act,\config')$ for $\act \in \globalActs$ (denoted as $\config \trans{\act} \config'$) that satisfy the following conditions:%
\begin{enumerate}
    \item for each $(\loc_i,\act_i!!,\loc_i')\in \act.\send$, $\accfac_{\act_i!!} \in \Nat_0$ is a \emph{send acceleration factor} that specifies how many processes take the given send transition,
    \item for each $\localfac = (\loc,\act??,\loc') \in \act.\recv$, $\accfac_{\localfac}$ is a \emph{receive acceleration factor} that specifies how many processes take the given receive transition,
    \item for all locations $\loc \in \Locs$,
          $$
              \config.\vk[\loc] = \sum\limits_{(\loc,\act_i!!,\loc') \in \act.\send}\! \accfac_{\act_i!!} + \sum\limits_{\localfac=(\loc,\act??,\loc') \in \act.\recv}\! \accfac_{\localfac},
          $$
          \ie the number of outgoing senders and outgoing receivers must sum up to the number of processes currently in $\loc$.
          Intuitively, this means that every process needs to execute either a send or receive action.
    \item $$\act.\low \leq \sum\limits_{0 \leq i < k} \accfac_{\act_i!!} \leq \act.\hi,$$ \ie the number of senders is inside the action's lower and upper bound,
    \item if \type is \tsender then $\accfac_{\act_i!!} \geq 1$ for $0 \leq i < k$, and
    \item if $\type$ is \tmaximal then $\!\!\!\sum\limits_{0 \leq i < k}\!\!\!\accfac_{\act_i!!} = \min{\act.\hi, \!\!\sum\limits_{(\loc,a_i!!,\loc') \in \act.\send} \!\!\!\!\!\!\!\! \config.\vk[\loc]} \wedge \! \sum\limits_{0 \leq i < k} \!\!\! \accfac_{\act_i!!} \geq 1$, i.e., sending transitions are taken by the maximum number of available processes within the upper bound $\act.\hi$. %
    \item for all $\loc \in \Locs$,
          \[
              \config'.\vk[\loc] =  \sum\limits_{(l',\act_i!!,l) \in \act.\send}\! \accfac_{\act_i!!} + \sum\limits_{\localfac=(\loc',\act??,\loc) \in \act.\recv}\! \accfac_{\localfac},
          \]
    \item $\config'.\vg\,{=}\,\config.\vg$ and $\config'.\vp\,{=}\,\config.\vp$, \ie shared variables and parameters are unchanged.
\end{enumerate}

For a given transition $\config \trans{\act} \config'$, we will in the following use $\Accfacs!$ to denote the set of send acceleration factors $\Accfacs! = \{ \accfac_{\act_i!!} \mid  (\loc_i,\act_i!!,\loc_i') \in \act.\send\}$ and analogously $\Accfacs? = \{ \accfac_{\localfac}  \mid \localfac \in \act.\recv \}$.

Such a transition on a synchronous action is also called a \emph{synchronous step}.
Analogously, we will write $\config \trans{\noact} \config'$ if there exists a rule $\rl \in \Rules$ and $\config \trans{\rl} \config'$ and refer to it as an \emph{asynchronous step}.

\begin{remark}[Undefined Receive Transitions]\label{rem:blocking}
    Note that the semantics is \emph{blocking} if a process is in a location that does not have a receive transition for the given action (and this process does not take a sending action). 
    This follows from the third condition, which requires all processes to take some local actions.
\end{remark}

\section{(Un)Decidability of Model Checking}\label{sec:complexity}
\label{sec:pmcp}

In this section, we examine the complexity of the PMCP for \TAS.
To this end, we first define different problem classes and recap existing complexity results for models closely related to \TAS. 
To characterize different classes of the PMCP, we use a variant of cardinality constraints~\cite{DelzannoSTZ12,EichlerJW25}.
\begin{definition}[Cardinality Constraint~\cite{DelzannoSTZ12,EichlerJW25}]
     A \emph{cardinality constraint} $\cardConstraint$ is a formula in the following grammar, with $\loc \in \Locs$ and $x \in \Nat$:
    \[
        \cardConstraint ::= \, \#\loc \geq x %
        \mid \#\loc = 0 %
        \mid \cardConstraint \land \cardConstraint %
        \mid \cardConstraint \lor \cardConstraint %
    \]
\end{definition}

The satisfaction of a cardinality constraint by a configuration $\config$ is defined in the natural way, \eg  $\config \models \#\loc \geq x$ iff $\config.\vk[\loc] \geq x$.

\begin{definition}[Cardinality Reachability Problem~\cite{DelzannoSTZ12,EichlerJW25}]
    Given a counter system of a threshold automaton (with synchronous actions) $\CS{\env, \syncta}$ and cardinality constraint $\varphi$, the \emph{cardinality reachability problem (CRP)} asks whether a configuration $\config$ with $\config \models \varphi$ is reachable in the system.
\end{definition}

\noindent
We denote by $CC[\geq x, =0]$ the class of all cardinality constraints, and by $CC[\geq x]$ the class of cardinality constraints in which atomic propositions are only of the form $\#\loc \geq x$.
In \cite{BaumeisterEJSV24}, specifications from $CC[\geq x]$ are called \emph{(general) parameterized coverability} specifications, and those from $CC[\geq x, =0]$ \emph{parameterized reachability} specifications.

Intuitively, a parameterized coverability specification specifies a set of locations and asks whether there exists a run that reaches a configuration such that in all the specified locations there is at least one process.
In a parameterized reachability constraint, one can additionally specify a set of locations that need to be empty in the target configuration.%

\smartpar{PMCP for TAs} 
For threshold automata (according to \cref{def:ta}), problems of the class $CC[\geq x, = 0]$ have been proven \NP-complete~\cite{Balasubramanian20,BaumeisterEJSV24}.

\subsection{Undecidability for Synchronous Actions}

We now investigate decidability of the CRP for \TAS.
As mentioned in \cref{rem:blocking}, in our semantics an action is blocked when there are processes that are not assigned to a send action and that cannot execute any local receive action. 
This blocking semantics can be seen as equivalent to the addition (see \cref{sec:blocking-guarded}) of a conjunctive guard, which is known to make safety problems undecidable for systems with synchronous (broadcast) steps~\cite{EmersonK03}.
It follows that problems of the class $CC[\geq x]$ are undecidable for \TAS under the blocking semantics.

A natural question is whether this undecidability result can be circumvented by relaxing the semantics to be non-blocking, or by restricting \TAS such that blocking cannot occur. 
In the following we show that undecidability even arises when blocking cannot occur (and therefore also under a non-blocking semantics).

For a given \TAS we say that receive actions are \emph{total} if every location $\loc$ has at least one outgoing rule with receive action $\act??$ for every action $\act \in \globalActs$.

\begin{theorem}\label{thm:undec-non-block}
    The CRP for $CC[\geq x]$ is undecidable for TA with synchronous actions, even if receive actions are total.
\end{theorem}
\begin{proof}[Sketch]
    We prove undecidability by reducing from the halting problem in two counter machines (2CM)~\cite{minsky}.
    For the reduction, we reuse the construction from \cite[Section 5]{EsparzaFM99}, which proved that liveness properties in broadcast protocols are undecidable.

    First note that an action in a broadcast protocol is equivalent to an \tsender action with upper and lower bound $1$. 
    Therefore, we can utilize the same construction, replacing broadcast actions with an equivalent \tsender action.
    By the definition of broadcast protocols in \cite{EsparzaFM99}, all receives are total. 

    The construction simulates state changes, increments and decrements of the 2CM faithfully by letting one process simulate the control state of the 2CM and all others simulate a unary encoding of the two counter values.
    Only zero tests are not simulated faithfully: instead of blocking when the given counter is non-zero, the simulation moves all processes representing a $1$ in the unary encoding of the counter to a sink location.

    Building on this idea, in a \TAS we can implement a check whether all zero tests along a run were faithful.
    Let the parameter $n$, and therefore $\Np = n$ denote the number of processes executing the system.
    We add a new \tmaximal action $\actdec{check}$ which on every location except the sink location, has a send and receive self-loop transition.
    The sink location only has a receive self-loop (making all receives total again).
    Additionally, we add an auxiliary location that can be reached from the halting location with local send action $\actdec{check}!!$.
    We set the upper and lower bound of the action $\actdec{check}$ to $\actdec{check}.\low = \actdec{check}.\hi = n$.
    
    Since an execution of $\actdec{check}$ requires all processes to be in a location besides the sink location, it can only be executed if no process ever moved to the sink location.
    Therefore, the auxiliary location can be reached iff the 2CM can reach a halting state.
    It follows that problems of the class $CC[\geq x]$ are undecidable.
    \qed
\end{proof}

\section{Recovering Decidability}\label{sec:reocver-dec}

We have shown that even a very restricted class of the CRP is undecidable in presence of synchronous actions.
In the following we introduce a bounded model checking (BMC) procedure that checks correctness up to a bounded number of synchronous steps.
Afterward, we introduce a class of \TAS with a static bound on the number of synchronous steps that can appear on any run, recovering decidability for this class.

\subsection{Bounded Model Checking}\label{sec:bmc}

To obtain a BMC procedure, we encode the reachability relation of a \TAS into SMT constraints.
We first recapitulate a similar approach for TA that we build upon.

\smartpar{Recap: Reachability in TA \cite{Balasubramanian20}}\label{sec:recap-bala}
The encoding presented in \cite{Balasubramanian20} relies on the observation that shared variables can only increase along a run of a TA.
Consequently, on any run of a TA, lower guards %
can only become enabled along an execution, and when enabled, they remain enabled forever.
Similarly, upper guards %
can only become disabled, and then remain disabled forever.

It follows that on any run of a TA with $k$ distinct guards, there can be at most $k$ steps where a guard changes its valuation.
Such a step is referred to as a \emph{context switch}, and the set of enabled guards is referred to as a \emph{context}.
Note that a context determines which rules are currently enabled in a run.

The SMT encoding then encodes all possible executions between context switches %
and combines it with an encoding of an explicit step where one guard changes its valuation.
The result is a formula $\phireachta(\config, \config')$ (called $\phireach$ in \cite{Balasubramanian20}) in Presburger arithmetic, which is true if and only if $\config'$ is reachable from $\config$ in threshold automaton $\ta$.

\smartpar{SMT Encoding for Synchronous Steps}
To encode a single synchronous step $\config \trans{\act} \config'$ with acceleration factors $\Accfacs!, \Accfacs?$, first let $\phiacc(\config, \config', \Accfacs!, \Accfacs?, \act)$ be a straightforward encoding of conditions $3, 4$ and $7$ from the definition of the semantics of \TAS (see \cref{sec:semantics}).  The full formula can be found in \cref{sec:complete-smt}.

Then a synchronous step of type \tsender, with arbitrary acceleration factors, additionally requires that each sending action is executed by at least one process. This can be encoded as
\begin{align*}
    \phisend(\config, \config',\act) & :=
    \exists \Accfacs!, \Accfacs?. \left( \bigwedge\limits_{\accfac \in \Accfacs!}\!\accfac \geq 1 \right) \wedge \phiacc(\config, \config', \Accfacs!, \Accfacs?, \act)
\end{align*}
\noindent
For steps of type \tmaximal, the maximum number of available senders has to execute the action.
We encode this as
\begin{align*}
    \phimax(\config, \config',\act)  := &
    \exists \Accfacs!, \Accfacs?. \phiacc(\config, \config', \Accfacs!, \Accfacs?, \act) \wedge \sum\Accfacs! \geq 1 \\
                                       & \wedge \left( \!\! \sum\! \Accfacs! = \act.\hi
    \vee  \left( \sum\!\Accfacs! = \!\!\! \sum\limits_{ \loc \in \{ \loc \mid (\loc,a_i!!,\loc') \in \act.\send \}} \!\!\!\!\!\!\! \config.\vk[\loc] \right) \right)
\end{align*}
\noindent
$\phisend$ and $\phimax$ evaluate to true for $(\config,\config',\act)$ if and only if $\config \trans{\act} \config'$ is a valid synchronous step of the given type.

Both formulas can be combined into $\phisync$, which intuitively allows selecting one action from all potential actions.
To distinguish which action has been executed and to ensure that at most one action is executed, we introduce another set of variables $S = \{ s_\act \, | \, \act \in \globalActs \}$.
\begin{align*}
    \phisync(\config, \config') := &\, \exists S. \,
    \sum\limits_{\act \in \globalActs} s_\act \leq 1     \wedge
    \!\!\bigwedge\limits_{\act \in \{ \act \in \globalActs \mid \act.\type = \tsender \}} \!\!\!\!\!\!\!\! (s_{\act} {=} 1 \Rightarrow \phisend(\config, \config', \act))
    \\ &\, \wedge \bigwedge\limits_{\act \in \{ \act \in \globalActs \mid \act.\type = \tmaximal \}}  \!\!\!\!\!\!\! (s_{\act} {=} 1 \Rightarrow \phimax(\config, \config', \act))
                                                                                    \\
                                  & \, \wedge \left( \sum\limits_{\act \in \globalActs} s_\act = 0 \Rightarrow \bigwedge\limits_{\loc\in\Locs} \config.\vk[\loc] = \config'.\vk[\loc] \right) \wedge \config.\vg = \config'.\vg \wedge  \config.\vp = \config'.\vp
\end{align*}
\noindent
$\phisync$ is constructed such that an action $\act$ is chosen iff $s_\act$ is set to one, and either $\phisend$ or $\phimax$ has to hold, depending on the type of $\act$. 
Note that $\phisync$ also allows no action to be selected, and then requires that processes remain in their locations.
This will become useful, as an encoding with $k$ instances of $\phisync$ can directly represent runs with $k$ \emph{or fewer} synchronous steps.

\smartpar{Combining Synchronous and Asynchronous Steps}
Now we combine our encoding for synchronous steps with the encoding for reachability of asynchronous steps.
Let $\syncta$ be a threshold automaton with synchronous actions and $k_\ta = k + 1$, where $k$ is the number of distinct guards that appear in $\syncta$. For a bound $k_\sync \in \mathbb{N}$ on the number of synchronous steps and for configurations $\config, \config'$ of $\syncta$, we define
\begin{align*}
    \phireach(k_\sync, \config, \config') := & \, \exists \config_0, \ldots, \config_{k_\sync}, \config_0^1, \ldots, \config_{k_\sync}^{k_\ta}. \, \config = \config_0 \wedge \config' = \config_{k_\sync}^{k_\ta} \\
                                            & \wedge \bigwedge\limits_{0 \leq i < k_\sync} \!\! \left( \phireachta (\config_i, \config_i^{k_\ta})
    \wedge \phisync(\config_i^{k_\ta},\config_{i+1}) \right)                                                                                                                                                      \\
                                            & \wedge \, \phireachta(\config_{k_\sync}, \config_{k_\sync}^{k_\ta})
\end{align*}
By construction, $\phireach$ evaluates to true if configuration $\config'$ is reachable from $\config$ within $k_\sync$ synchronous steps.
Note that $\phireach$ allows for up to $k_\ta$ context switches between every synchronous step.
This is because without additional arguments, it is not possible to infer how context switches interleave with synchronous steps, see \cref{sec:context-changes-sync-ta} for more details.
A more formal correctness argument for the SMT encoding can be found in \cref{sec:correct-smt}.
Note that $\phireach$ is still a formula in existential Presburger arithmetic.

\subsection{Bounded Synchronous Steps}

While bounded model checking can be useful in practice, it cannot certify correctness.
To address this limitation, we introduce a class of \TAS for which a static bound on the number of synchronous steps across all runs exists.
This class is inspired by the observation that modules often represent isolated steps that are executed by all processes before the system can make further progress.
This is, for example, the case for the weak consensus module within graded consensus (\cref{ex:weak-cons-ta-intro,ex:graded-cons,ex:weak-cons-sync}).
We therefore partition the set of locations into \emph{layers}, such that every synchronous step moves all processes into the next layer.

\begin{definition}[Layered \TAS]\label{def:layered}
    A \TAS $\syncta$ is \emph{layered} if there exists a partition $\Part = \{\layer_1, \dots, \layer_m\}$ or $\Part = \{\layer_1, \dots, \layer_m, \sinkLayer\}$ of the set of locations $\Locs$, and for every $\layer_i$,
    \begin{itemize}
        \item local rules cannot transition to a lower layer, \ie for every $ \loc \in \layer_i, \loc'\in \Locs$ s.t. $\loc \trans{\noact} \loc'$, it holds that $\loc' \in \bigcup\limits_{j \in i,\dots,m} \layer_j \cup \sinkLayer$.
        \item every global rule must transition to a higher layer or $\layer_\sink$, \ie for every $\loc \in \layer_i, \loc'\in \Locs$ s.t. $\loc \trans{\someact} \loc'$, it must hold that $\loc' \in \bigcup\limits_{j \in i + 1,\dots,m} \!\! \layer_j \cup \sinkLayer$.
        \item for all locations in $\sinkLayer$ (if any), all outgoing transitions are synchronous receive self-loops, \ie for all $\loc \in \sinkLayer, \loc' \in \Locs$ with $\loc \trans{\someact} \loc'$ it holds that $\loc = \loc'$ and $\someact = \act??$ for some $\act \in \globalActs$. 
            We also call $\sinkLayer$ the \emph{sink layer}.
    \end{itemize}
\end{definition}

We define the \emph{number of layers} as $|\Part|$.
Regular TA are by definition layered, with a single layer.

Note that \cref{def:layered} requires all local receive actions to transition to a higher layer or the sink layer, which implies that receive self-loops are only allowed in the sink layer.
We will show in \cref{sec:eval} that many synchronous abstractions are layered.

We can automatically check whether a given \TAS is layered, and compute the layers if it is. The following result provides the theoretical basis of the algorithm, more details and a proof of the lemma are provided in \cref{sec:layering-alg}. 

\begin{restatable}{lemma}{lemLayeredScc}\label{lem:layered-scc}
    A \TAS $\syncta$ is layered iff there does not exist a synchronous transition within any SCC of the underlying graph, except for the synchronous receive self-loops in sink locations.
\end{restatable}

Observing that every synchronous step moves \emph{all} processes either to a strictly higher layer, or to $\sinkLayer$, and that locations in $\sinkLayer$ do not have send actions, we get a bound on the number of synchronous steps in any run of a layered \TAS.

\begin{lemma}\label{thm:layerd-sync-bounded}
    Let $\syncta$ be a layered \TAS with $|\Part|=m$. Then every run of $\syncta$ has at most $m-1$ synchronous steps.
\end{lemma}

Since the number of synchronous steps in a layered \TAS is bounded, our BMC procedure instantiated with the proper bound is a decision procedure for the CRP.

\begin{theorem}
    Problems in $CC[\geq x, =0]$ for a layered \TAS are decidable in NP.
\end{theorem}
\begin{proof}[Sketch]
    Decidability follows from \cref{thm:layerd-sync-bounded} and correctness of $\phireach$.
    The complexity bound follows from the fact that $\phireach$ is a polynomial-size formula in existential Presburger arithmetic which can be checked in \NP~\presbCite.
\end{proof}

\section{Modular Models \& Replacement}\label{sec:modular-replace}

The idea of synchronous module abstraction is to reduce the complexity by replacing modules with a concise synchronous abstraction.
Before applying this technique we first formally define the notion of replacement.

\subsection{Replacement Operations}

In the following, we fix an environment $\env$ in which all automata are executed.
Then for a \TAS a \emph{replacement operation} $\replacement$ is defined by
\begin{itemize}[topsep=2pt, itemsep=0pt]
    \item $\LocsRM \subseteq \Locs$, the set of locations to remove from the automaton,
    \item  $\LocsAdd$ the set of locations to add to the automaton,
    \item the set of rules $\RulesAdd$ and global rules $\globalRulesAdd$ to add,
    \item and the set of shared variables to remove\footnote{Since none of our examples requires it, we do not allow the addition of shared variables. This could be supported at the cost of notational complexity.} $\VarsRM$.
\end{itemize}

For a replacement $\replacement$, we define the set of \emph{entry locations $\LocsIn$} as $\LocsIn = \{ \loc \in \Locs \setminus \LocsRM \mid \exists \loc' \in \LocsRM. \, \loc \trans{\noact} \loc' \text{ or } \loc \trans{\someact} \loc' \}$.
Intuitively, it contains the locations with transitions into the replaced locations.
Analogously, we define the \emph{exit locations $\LocsOut = \{ \loc' \in \Locs \setminus \LocsRM \mid \exists \loc \in \LocsRM. \, \loc \trans{\noact} \loc' \text{ or }  \loc \trans{\someact} \loc' \}$}, which describe locations with transitions out of the replaced locations.

To be well-defined, variables in $\VarsRM$ cannot be referenced by any rule kept by the replacement, and initial locations cannot be removed, \ie $\initial{\Locs} \cap \LocsRM = \emptyset$.

\begin{definition}\label{def:tas-replace}
    For a \TAS $\syncta = \langle \Locs, \initial{\Locs}, \SharedVars,\Rules, \globalRules \rangle$ and replacement
    $\replacement = \langle \LocsRM,\allowbreak \LocsAdd, \RulesAdd, \globalRulesAdd, \VarsRM \rangle$, the resulting \TAS after applying replacement $\replacement$, denoted as $\replace{\syncta}{\replacement} = \langle \Locs', \initial{\Locs}', \SharedVars', \Rules', \globalRules' \rangle$ is defined as
    \begin{itemize}[topsep=2pt, itemsep=0pt]
        \item $\Locs' = (\Locs \setminus \LocsRM ) \cup \LocsAdd$ and $\initial{\Locs}' = \initial{\Locs}$
        \item $\SharedVars' = \SharedVars \setminus \VarsRM$
        \item  $\Rules ' =  \RulesAdd \cup \{ \rl \in \Rules \mid \rl.\vfrom \notin \LocsRM \wedge \rl.\vto \notin \LocsRM \}$, and
        \item $\globalRules' =  \globalRulesAdd \cup  \{ (\act,\_,\_,\_,\_,\_) \in \globalRules \mid \forall (\loc,\_, \loc') \in \act.\send \cup \act.\recv. \, \loc, \loc' \notin \LocsRM \}$.%
    \end{itemize}
\end{definition}
\begin{example}\label{ex:replace-weak}
    \begin{figure}
    \begin{subfigure}[c][34mm]{0.43\textwidth}
        \centering
        \scalebox{0.7}{
            \begin{tikzpicture}[lts]
                \draw [cispa-anthracite, dashed, semithick, rounded corners, fill=cispa-dark-gray, fill opacity=0.06] (-0.45, 1.8) rectangle (3.2, -1.8);
                \node [anchor=south west, text=cispa-anthracite] at (-0.45, 1.8) {$\ta_{WC}$};

                \draw [cispa-orange, dash dot, semithick, rounded corners, fill=cispa-yellow, fill opacity=0.1] (2.4, 1.6) rectangle (6.35, -1.6);
                \node [anchor=south east, text=cispa-orange] at (6.35, 1.6) {$A_{GC}$};

                \node (i0) [location, init, draw=cispa-red, dashed, thick] at (0, 0.75) {$x_0$};
                \node (i1) [location, init, draw=cispa-red, dashed, thick] at (0, -0.75) {$x_1$};

                \node (cnt) [other-location] at (1.4, 0) {$\loc_{cnt}$};

                \node (o0) [location, draw=cispa-green, densely dotted, thick] at (2.8, 1.25) {$y_0$};
                \node (of) [location, draw=cispa-green, densely dotted, thick] at (2.8, 0) {$y_\bot$};
                \node (o1) [location, draw=cispa-green, densely dotted, thick] at (2.8, -1.25) {$y_1$};

                \node (gcnt) [location] at (4.3, 0) {$\loc_{cnt}$};

                \node (g00) [location] at (5.7, 1.2) {$\loc_{0,0}$};
                \node (g01) [location] at (5.7, 0.4) {$\loc_{0,1}$};
                \node (g10) [location] at (5.7, -0.4) {$\loc_{1,0}$};
                \node (g11) [location] at (5.7, -1.2) {$\loc_{1,1}$};

                \path (0, -1.9);

                \path[->]
                    (i0) edge [draw=cispa-red, thick] node [above, sloped, yshift=-2pt] {...} (cnt)
                    (i1) edge [draw=cispa-red, thick] node [below, sloped, yshift=1.5pt] {...} (cnt)

                    (cnt) edge [draw=cispa-green, thick] node [above, sloped, yshift=-2pt] {...} (o0)
                    (cnt) edge [draw=cispa-green, thick] node [above, yshift=-2pt] {...} (of)
                    (cnt) edge [draw=cispa-green, thick] node [below, sloped, yshift=1.5pt] {...} (o1)

                    (o0) edge [] node [above, sloped, yshift=-2pt] {...} (gcnt)
                    (of) edge [] node [above, yshift=-2pt] {...} (gcnt)
                    (o1) edge [] node [below, sloped, yshift=1.5pt] {...} (gcnt)

                    (gcnt) edge [] node [above, sloped, yshift=-2pt] {...} (g00)
                    (gcnt) edge [] node [above, sloped, yshift=-2pt] {...} (g01)
                    (gcnt) edge [] node [below, sloped, yshift=1.5pt] {...} (g10)
                    (gcnt) edge [] node [below, sloped, yshift=1.5pt] {...} (g11)
                ;
            \end{tikzpicture}
        }
        \caption{Full graded consensus as TA. Locations in \textcolor{gray}{$\LocsRM$} are depicted in \textcolor{gray}{gray}.}
        \label{fig:replace-graded-before}
    \end{subfigure}
    \hfill
    \begin{subfigure}[c][34mm]{0.54\textwidth}
        \centering
        \scalebox{0.7}{
            \begin{tikzpicture}[lts]
                \draw [cispa-dark-blue, dashed, semithick, rounded corners, fill=cispa-dark-blue, fill opacity=0.06] (-0.45, 1.8) rectangle (5.2, -1.8);
                \node [anchor=south west, text=cispa-dark-blue] at (-0.45, 1.8) {$\syncta_{WC}$};

                \draw [cispa-orange, dash dot, semithick, rounded corners, fill=cispa-yellow, fill opacity=0.1] (4.35, 1.6) rectangle (8.3, -1.6);
                \node [anchor=south east, text=cispa-orange] at (8.3, 1.6) {$\ta_{GC}$};

                \node (i0) [location, init, draw=cispa-red, dashed, thick] at (0, 0.75) {$x_0$};
                \node (i1) [location, init, draw=cispa-red, dashed, thick] at (0, -0.75) {$x_1$};

                \node (o0) [location, draw=cispa-green, densely dotted, thick] at (4.75, 1.25) {$y_0$};
                \node (of) [location, draw=cispa-green, densely dotted, thick] at (4.75, 0) {$y_\bot$};
                \node (o1) [location, draw=cispa-green, densely dotted, thick] at (4.75, -1.25) {$y_1$};

                \node (gcnt) [location] at (6.25, 0) {$\loc_{cnt}$};

                \node (g00) [location] at (7.65, 1.2) {$\loc_{0,0}$};
                \node (g01) [location] at (7.65, 0.4) {$\loc_{0,1}$};
                \node (g10) [location] at (7.65, -0.4) {$\loc_{1,0}$};
                \node (g11) [location] at (7.65, -1.2) {$\loc_{1,1}$};

                \path (0, -1.9);

                \path[->, nodes={font=\scriptsize}]
                    (i0) edge [draw=blue, thick, bend left=15] node [above, sloped, yshift=-2pt] {$\actdec{z}!!,\actdec{mz}!!,\actdec{z}??$} (o0)
                    (i1) edge [draw=blue, thick, bend right=15] node [below, sloped, yshift=1.5pt] {$\actdec{o}!!,\actdec{mo}!!,\actdec{o}??$} (o1)

                    (i1) edge [draw=blue, thick, bend left=12] node [sloped, xshift=12pt, yshift=-2pt] {$\actdec{z}??,\actdec{mz}??$} (o0)
                    (i0) edge [draw=blue, thick, bend right=12] node [below, sloped, xshift=12pt, yshift=1.5pt] {$\actdec{o}??,\actdec{mo}??$} (o1)

                    (i0) edge [draw=blue, thick] node [above, sloped, xshift=24pt, yshift=-2pt] {$\actdec{mz}!!,\actdec{b}!!,\actdec{mo}??$} (of)
                    (i1) edge [draw=blue, thick] node [below, sloped, xshift=24pt, yshift=1.5pt] {$\actdec{mo}!!,\actdec{mz}??,\actdec{b}??$} (of)
                ;

                \path[->]
                    (o0) edge [] node [above, sloped, yshift=-2pt] {...} (gcnt)
                    (of) edge [] node [above, yshift=-2pt] {...} (gcnt)
                    (o1) edge [] node [below, sloped, yshift=1.5pt] {...} (gcnt)

                    (gcnt) edge [] node [above, sloped, yshift=-2pt] {...} (g00)
                    (gcnt) edge [] node [above, sloped, yshift=-2pt] {...} (g01)
                    (gcnt) edge [] node [below, sloped, yshift=1.5pt] {...} (g10)
                    (gcnt) edge [] node [below, sloped, yshift=1.5pt] {...} (g11)
                ;
            \end{tikzpicture}
        }
        \caption{Graded consensus after replacement. New transitions are depicted in \textcolor{blue}{blue}.}
        \label{fig:replace-graded-after}
    \end{subfigure}
    \caption{Replacement of weak consensus ($\ta_{WC}$, \cref{fig:weak-cons-ta}) in graded consensus ($\ta_{GC}$, \cref{fig:graded-cons-ta}) by its synchronous model ($\syncta_{WC}$, \cref{fig:weak-cons-sync}). 
    $\globalRulesAdd$ marked in \textcolor{blue}{blue}, $\LocsIn$ and transitions into the module in \textcolor{cispa-red}{red}, $\LocsOut$ and transitions out of the module \textcolor{cispa-green}{green}.}
    \label{fig:replace-graded}
\end{figure}

    \cref{fig:replace-graded} depicts the replacement of weak consensus by its synchronous abstraction in graded consensus (from \cref{ex:graded-cons}).
    The replacement removes the shared variables $\VarsRM = \{ c_0, c_1, v \}$, the counting location $\LocsRM = \{ \loc_{cnt} \}$, and $5$ local rules, all of which are internal to the weak consensus TA $A_{WC}$.
    It adds the $5$ global rules from $\syncta_{WC}$, but no locations or local rules.
\end{example}

\subsection{Faithful Replacements}

Replacements allow us to create automata with entirely different behavior than the original automaton.
To be practical for model checking, a replacement needs to preserve properties of interest.
To this end, we introduce the notion of a reachability-preserving replacement:
\begin{definition}[Reachability-Preserving Replacement]\label{def:reach-preserve}
    For \TAS $\syncta$, a replacement operation $\replacement = \langle \LocsRM, \LocsAdd, \RulesAdd, \globalRulesAdd, \VarsRM \rangle$ is \emph{reachability-preserving (rp)} if for every reachable configuration $\config$ with $\forall \loc \in \LocsRM. \, \allowbreak \config.\vk[\loc] = 0$, there exists a reachable configuration $\config'$ in $\replace{\syncta}{\replacement}$ with $\config.\vp = \config'.\vp, \forall \var \in \SharedVars \setminus \VarsRM. \, \config.\vg[\var] = \config'.\vg[\var]$, $\forall \loc \in \Locs \setminus \LocsRM. \,\config.\vk[\loc]  = \config'.\vk[\loc]$ and $\forall \loc \in \LocsAdd. \, \config'.\vk[\loc] = 0$.
    It is \emph{strongly reachability-preserving (srp)} if the other direction holds too.
\end{definition}

Intuitively, \cref{def:reach-preserve} requires that for any reachable configuration where no process is in $\LocsRM$, an equivalent configuration must be reachable after the replacement.
It follows that if a configuration satisfying $\varphi \wedge \bigwedge_{\loc \in \LocsRM} \#\loc = 0$, for some cardinality constraint $\varphi$, is reachable in $\syncta$,
    then a configuration satisfying $\varphi \wedge \bigwedge_{\loc \in \LocsAdd} \#\loc = 0$ is reachable in
    $\replace{\syncta}{\replacement}$.

    The restriction that locations $\LocsRM$ must be empty is necessary since a synchronous step is executed by all processes, whereas asynchronous steps move one process at a time.
Therefore, asynchronous steps create many intermediate configurations, which the synchronous action abstracts from.

Note that a reachability-preserving replacement can add new behavior, \ie
it allows $\replace{\syncta}{\replacement}$ to be an overapproximation of $\syncta$.
Overapproximations can be easier to check, as we will see in \cref{sec:eval}, and if an overapproximation is safe, the original system is also guaranteed to be safe.
However, reported counterexamples might not be valid behavior of $\syncta$.
The stronger notion of \emph{strongly reachability preserving-replacement} prevents such spurious counterexamples.

Whether a replacement is reachability preserving can again be expressed by an SMT formula.

\begin{theorem}\label{thm:reach}
    Let $\syncta$ and $\replace{\syncta}{\replacement}$ be layered \TAS with $k_\sync = |\Part_\syncta|{-}1$ and $k'_\sync=|\Part_{\replace{\syncta}{\replacement}}|{-}1$. Then $\replacement$ is reachability-preserving iff  the following formula holds:
    \begin{align*}
        \phieq(\syncta, \replacement) := \, & \forall \initial{\config}, \config \in \Configs_{\syncta}.\, \bigg( \phiinit(\initial{\config}) \wedge \phireach(k_\sync, \initial{\config}, \config) \wedge \bigwedge\limits_{\LocsRM}^\loc \config.\vk[\loc] = 0         \\
                                            & \, \Rightarrow \exists \initial{\config'},\config'\in \Configs_{\replace{\syncta}{\replacement}}.\, \phiinit(\initial{\config'}) \wedge \phireach(k'_\sync, \initial{\config'},\config') \wedge \config.\vp = \config'.\vp \\
                                            & \, \wedge \!\bigwedge\limits_{\Locs \setminus \LocsRM}^\loc  \!\! \config.\vk[\loc]  = \config'.\vk[\loc]
        \wedge \!\bigwedge\limits_{\LocsAdd}^\loc \config'.\vk[\loc] = 0 \wedge  \!\bigwedge\limits_{\SharedVars \setminus \VarsRM}^{\var} \!\! \config.\vg[\var] = \config'.\vg[\var] \bigg)
    \end{align*}
\end{theorem}

Here, $\phiinit(\config)$ denotes the formula that evaluates to true iff $\config$ is an initial configuration.
The details of $\phiinit$ and an encoding for strongly reachability-preserving can be found in \cref{sec:smt-reach-preserve}.
If $\syncta$ or $\replace{\syncta}{\replacement}$ are not layered, $\phieq$ can only check that the replacement is reachability-preserving for a bounded length of runs.

Note that the SMT formula $\phieq$ contains a quantifier alternation.
Deciding the validity of such a formula is \coNEXP-complete~\cite{Haase14}
.\footnote{
    $\phieq$ contains quantifier alternation $\forall\exists$ (with no free variables).
    The formula is therefore in $\Pi_2$ and its negation in $\Sigma_2$.
    Since deciding a formula in $\Sigma_2$ is \NEXP-complete \cite[Thm. 1]{Haase14}, it follows that deciding formulas in $\Pi_2$ is \coNEXP-complete.
}

In the following we will present techniques and conditions that enable a practical use of replacements despite this worst-case complexity, and \cref{sec:eval} will show that our approach is efficient on practical examples.

\subsection{Practical Synchronous Module Abstraction}
Formula $\phieq$ contains a quantifier alternation and encodes the full reachability relations of $\syncta$ and $\replace{\syncta}{\replacement}$, making it inefficient to check if used naively.
However, under the following conditions the checks can be done \emph{locally} on the replaced part of the automaton:
\begin{enumerate*}[label=(\roman*)]
\item $\LocsIn = \initial{\Locs}$,
\item processes can pass through the replacement only once,
\item $\replacement$ does not modify shared variables used in the rest of $\syncta$, and 
\item $\globalRulesAdd \cap \globalRules_\syncta = \emptyset$.
\end{enumerate*}
More details can be found in \cref{sec:local-replace-cond}.

When checking $\phieq$ locally, the required reachability checks can be much simpler, and the quantifier alternation can effectively disappear: if $\LocsAdd = \RulesAdd = \emptyset$, then the exit locations will always be reached after a single synchronous step.
In this case, the values of all existentially quantified configurations in $\phieq$ are fixed by the universally quantified configurations, and $\phireach(k'_\sync, \initial{\config'},\config')$ becomes a finite disjunction over $\globalRulesAdd$.  

While these conditions are a significant restriction, they are satisfied in \cref{ex:replace-weak} and most of the cases considered in \cref{sec:eval}.
This is not surprising, as a standard use of such modules is to replace a part of $\syncta$ by something that amounts to a function call in the algorithm, and that can be modeled as a single synchronous step.

\section{Experimental Evaluation}\label{sec:eval}

\smartpar{Applying Synchronous Module Abstraction}
To evaluate our technique, we study the king consensus algorithm \cite{BermanGP89} as presented in \cite{FitziM00}.
The algorithm is split into three modules: weak consensus (\cref{alg:weak-cons}), graded consensus (\cref{alg:graded-cons}), and lastly king consensus (presented in detail in \cref{sec:missing-models}).

To verify the algorithm, we start at the lowest layer, the weak consensus algorithm, and construct a TA $A_{WC}$ 
(\cref{fig:weak-cons-ta}). 
After verifying its desired properties (using the procedure from \cref{sec:reocver-dec}), instead of directly proceeding with the next module, we construct a synchronous abstraction $\syncta_{WC}$.
We verify that $\syncta_{WC}$ constitutes a faithful replacement of $A_{WC}$ (via \cref{thm:reach}),
ensuring that $\syncta_{WC}$ is an overapproximation of $A_{WC}$ and can be used as a module when constructing the automaton for graded consensus.
Optionally, we can check whether $\syncta_{WC}$ is strongly reachability preserving to avoid spurious counterexamples later. In this case, it is---if it was not, we could refine the model further.

We continue this approach, next using $\syncta_{WC}$ as a module in the definition of graded consensus, verify its properties and obtaining a synchronous abstraction that is in turn used as a module in the definition of king consensus.
We will see in the following that this synchronous module abstraction technique leads to much smaller models, which additionally can be checked much quicker.

\smartpar{Models} 
For our evaluation, we also construct the full TA models for each algorithm.
All automata are encoded in the file format of ByMC~\cite{0001W18}, extended with constructs for synchronous actions: $A_{WC}$ (\cref{fig:weak-cons-ta}) is encoded in \texttt{weak.ta}, $\syncta_{WC}$ (\cref{fig:weak-cons-sync}) in \texttt{weak.sta}, and the combination of $\syncta_{WC}$ and $A_{GC}$ (\cref{fig:replace-graded-after}) in \texttt{graded-wc.sta}.
For graded consensus, we construct two different synchronous abstractions, 
an overapproximation (\texttt{graded-over.sta}) and an exact representation (\texttt{graded-exact.sta}). 
Additional details can be found in \cref{sec:missing-models}.

\begin{figure}
    \centering
    \begin{minipage}[c][39mm]{0.48\textwidth}
        \centering
\captionof{table}{Model size  for synchronous and asynchronous  algorithm models. 
``over'' denotes an overapproximation and ``exact'' an exact representation of an algorithm.
}
\label{tab:m-size}
\scalebox{0.84}{
    \begin{tabular}{@{}l | c c | c c @{}}
        \toprule
         & \multicolumn{2}{c|}{TA} & \multicolumn{2}{c}{TAS} \\
        Algorithm          & \, $|\Locs|$ \, & \, $|\Rules|$ \, & \, $|\Locs|$\, &  \,$|\globalRules|$\, \\
        \midrule
        Weak & 6 & 5 & 5 & 5 \\
        Graded (over)& 11 & 12 & 6 & 5 \\
        Graded (exact) & 11 & 12 & 7 & 7 \\
        King  & 13 & 22 & 4 & 3 \\
        \bottomrule
    \end{tabular}
}

    \end{minipage}
    \hfill\hfill
    \begin{minipage}[c][39mm]{0.49\textwidth}
        \centering
\captionof{table}{
    Model Checking on \TAS (\texttt{.sta}), TA (\texttt{.ta}), timeout 30min}
\label{tab:mc}
\scalebox{0.84}{
    \begin{tabular}{@{}l rr@{}}
        \toprule
        Model                      & Time [s] & Mem.\ [MiB] \\
        \midrule
        \texttt{weak.ta}           & 0.17     & 37.2        \\
        \texttt{weak.sta}          & 0.06     & 34.1        \\
        \texttt{graded.ta}         & 8.73     & 80.9        \\
        \texttt{graded-wc.sta}     & 2.49     & 62.2        \\
        \texttt{graded-over.sta}   & 0.04     & 35.0        \\
        \texttt{graded-exact.sta}  & 0.07     & 35.4        \\
        \texttt{king.ta}           & 41.39    & 204.6       \\
        \texttt{king-gc-over.sta}  & 0.46     & 50.5        \\
        \texttt{king-gc-exact.sta} & 1.23     & 63.8        \\
        \texttt{king.sta}          & 0.03     & 34.6        \\
        \bottomrule
    \end{tabular}
}

    \end{minipage}
\end{figure}

\smartpar{Model Complexity} 
We compare the size of the respective models in \cref{tab:m-size}. 
We observe that the synchronous abstractions (in column \TAS) have significantly fewer locations, \eg king consensus shrinks from $13$ locations to just $4$.
Also the number of synchronous actions required is usually small and all synchronous models are layered.
In summary, the synchronous abstractions are far less complex than their asynchronous counterparts.

\smartpar{Model Checking Performance} 
We implemented a prototype of the decision procedure presented in \cref{sec:bmc} and checked correctness properties of every algorithm, encoded as reachability constraints. 
As an example, for weak consensus we check weak consistency: if some honest process accepts $0$ or $1$, other honest processes can only accept the same value or move to $y_\bot$, corresponding to unreachability of constraint $\varphi = y_0 \geq 1 \wedge y_1 \geq 1$.

\cref{tab:mc} presents the runtime results of our prototype (see \cref{sec:bench-setup} for more details on the setup). 
Overall, checking synchronous modules, or models based on synchronous abstraction, is significantly faster than checking the full TA.
For example, the king consensus TA is checked in $~41$s compared to just $0.03$s on the fully synchronous version, and $1.2$s when the graded consensus module is replaced by an exact synchronous abstraction. 
We can also see that using an overapproximation (that still satisfies all correctness properties) can further improve the performance dropping the runtime from $1.2$s for \texttt{king-gc-exact} to $0.5$s on \texttt{king-gc-over}.

\begin{table}
    \caption{Runtime for reachability-preserving checks with timeout of 30min}
    \label{tab:refine}
    \centering
    \scalebox{0.84}{
        \begin{tabular}{@{}ll | ccc| ccc@{}}
            \toprule
                                       &                            & \multicolumn{3}{c}{Reachability-preserving} & \multicolumn{3}{|c}{Strongly reachability-pres.}                                                                      \\
            \cmidrule(lr){3-5} \cmidrule(l){6-8}
            Original                   & Replacement                & holds                                       & Time [s]                                         & Mem.\ [MiB] & holds                       & Time [s] & Mem.\ [MiB] \\
            \midrule
            \texttt{weak.ta}           & \texttt{weak.sta}          & \cmark                                      & 0.23                                             & 43.4        & \cmark                      & 0.72     & 45.1        \\
            \texttt{graded.ta}         & \texttt{graded-wc.sta}     & \cmark                                      & 0.35                                             & 45.4        & \cmark                      & 1.00     & 46.0        \\
            \texttt{graded.ta}         & \texttt{graded-over.sta}   & \cmark                                      & 11.25                                            & 99.2        & \xmark                      & 18.54    & 103.5       \\
            \texttt{graded.ta}         & \texttt{graded-exact.sta}  & \cmark                                      & 12.39                                            & 108.4       & \cmark                      & 61.65    & 147.9       \\
            \texttt{graded-wc.sta}     & \texttt{graded-over.sta}   & \cmark                                      & 2.50                                             & 69.6        & \xmark                      & 5.19     & 77.8        \\
            \texttt{graded-wc.sta}     & \texttt{graded-exact.sta}  & \cmark                                      & 2.06                                             & 68.6        & \cmark                      & 7.91     & 102.4       \\
            \texttt{king.ta}           & \texttt{king-gc-over.sta}  & \cmark                                      & 10.77                                            & 98.0        & \xmark                      & 18.61    & 98.5        \\
            \texttt{king.ta}           & \texttt{king-gc-exact.sta} & \cmark                                      & 13.40                                            & 113.5       & \cmark                      & 66.24    & 159.7       \\
            \texttt{king.ta}           & \texttt{king.sta}          & \cmark                                      & 52.49                                            & 307.9       & \multicolumn{3}{c}{timeout}                          \\
            \texttt{king-gc-over.sta}  & \texttt{king.sta}          & \cmark                                      & 0.67                                             & 56.6        & \cmark                      & 2.59     & 65.4        \\
            \texttt{king-gc-exact.sta} & \texttt{king.sta}          & \cmark                                      & 3.04                                             & 79.7        & \cmark                      & 5.76     & 95.8        \\
            \bottomrule
        \end{tabular}
    }
\end{table}

\smartpar{Reachability-Preserving Checks}
\cref{tab:refine} presents the runtime results for verifying that the synchronous abstractions are faithful. 
All reachability-preserving checks terminate within one minute, most cases in under $15$s.
Therefore, we argue that these checks even with their theoretical worst case complexity, are quite efficient in practice.

Although not necessary for checking our safety properties, we also evaluate performance on checking whether a replacement is strong reachability-preserving takes significantly longer. 
We see that this takes significantly longer, with one case running into a timeout. 
Furthermore, we note that the benefits of our synchronous module abstraction also show in the reachability-preserving checks: while checking directly whether \texttt{king.sta} is an exact abstraction of \texttt{king.ta} times out, the checks from \texttt{king.ta} to \texttt{king-gc-exact.sta} and from \texttt{king-gc-exact.sta} to \texttt{king.sta} succeed, implying that \texttt{king.sta} is an exact abstraction of \texttt{king.ta}.
Finally, note that the models that are overapproximations are not strong reachability-preserving, as expected.

\smartpar{Summary}
We have seen that our synchronous modular abstraction technique can drastically improve model checking performance while also reducing the complexity of the individual models. 
We argue that our technique therefore makes parameterized model checking significantly more practical, even though the synchronous abstraction currently needs to be designed manually.

\section{Related Work}

\smartpar{Parameterized Model Checking}
The fully automatic verification of concurrent system models with a parametric number of processes is a long-standing and active area of research~\cite{Bloem2015,AbdullaST18,Esparza14}.
Many approaches use automata-based models, with different types of communication such as pairwise synchronization~\cite{GermanS92}, broadcast~\cite{EmersonN98,EsparzaFM99}, or guarded transitions~\cite{EmersonK00,EmersonK03}.
An inspiration for our synchronous actions are global synchronization protocols, which combine a generalization of broadcast synchronization with guarded transitions~\cite{JaberJW0S20,JaberWJKS21}.
However, like the other models mentioned so far, GSPs do not support the modeling of Byzantine fault-tolerant algorithms. 

\smartpar{Threshold Automata}
An important system model that directly supports modeling and verification of Byzantine fault-tolerant algorithms are threshold automata~\cite{JohnKSVW13,fmsdKonnovLVW17,KonnovLVW17,Kukovec0W18}.
Notably, threshold automata have also received significant tool support, first with the model checker ByMC~\cite{0001W18,0001LSW23}, then an SMT-based implementation by Balasubramanian et al. that lifted some restrictions of ByMC~\cite{Balasubramanian20}, and recently with the model checker TACO~\cite{EichlerBSDVJ26} that re-implements existing techniques along with new extensions of the TA model~\cite{BaumeisterEJSV24}.

The SMT-based approach of Balasubramanian et al. is of particular interest to us, as the technique we present is also based on an SMT encoding, re-uses some of their constraints and extends the approach to support synchronous actions.

Synchronous steps in TA have been considered before in the model of synchronous threshold automata (STA)~\cite{StoilkovskaKWZ19,StoilkovskaKWZ22,StoilkovskaKWZ21}, which are TA in which processes execute \emph{all} transitions in lockstep. 
In contrast, our model allows to mix synchronous and asynchronous steps, which is essential for the modular verification approach we present.

Another model derived from TA and close to ours are \emph{layered automata (LTA)} and their abstraction called \emph{guard automata~\cite{0001TW21}}.
LTA require a user to divide an algorithm into layers and require each transition to move a process into the next layer.
For cyclic LTA, which are LTA that repeat a finite sequence of layers forever, so-called guard automata can be constructed as overapproximations. 
Thus, in contrast to our approach, the framework does support algorithms over multiple rounds. 
However, the abstraction is not complete, therefore it might report spurious counterexamples, and similar to STA, communication via actions is not supported.
The tool PyLTA~\cite{ThomasS23} implements a decision procedure for LTL properties on guard automata.

\section{Conclusion}

In this work we have introduced threshold automata with synchronous steps, a computational model that combines asynchronous actions from threshold automata with expressive global synchronous actions.
While the addition of synchronous steps results in an undecidable PMCP in general, we introduced a bounded model checking procedure.
Moreover, we identified a subclass of automata for which a static completeness bound can be computed, making the PMCP decidable for this subclass.

The addition of synchronous actions enables our new technique of synchronous module abstraction:
instead of combining modular algorithm descriptions into complex monolithic models, our technique allows the designer to replace modules by concise synchronous abstractions.
We implemented our techniques in a prototype, and demonstrated its effectiveness on the example of the king consensus.
Using synchronous module abstraction, we reduced the model complexity and the model checking time significantly.
We believe that this technique can be even more effective on larger, more complex distributed algorithms.
Therefore, in the future we plan to evaluate our techniques systematically on more complex algorithms, closer to real-world distributed systems.

In addition, we plan to extend our line of work in two directions: 
we will investigate the automatic synthesis of the abstractions from the module description and its correctness properties, and we hope to make progress towards supporting multi-round algorithms, potentially by utilizing similar overapproximations as the line of work on guard automata~\cite{0001TW21}.

\newpage

\begin{credits}
    \subsubsection{AI Usage}
    AI has been used in the creation of this work.
    Specifically, it has been used to review drafts of this paper and in the prototype implementation, for example, to generate test cases from natural language inputs, but also to implement a GUI which we consider supplementary to our artifact.
    More details will be provided in the artifact.
    Models that have been used:
    Deepseek-v4-flash, Deepseek-v4.1-flash, GLM-5.3-flash, GLM-5.3, hosted on Ollama cloud; Fable 5.1, Opus 5.5 provided by Anthropic.

    \subsubsection{\ackname}
    We thank Julian Loss for initial discussions on modularly defined distributed algorithms.
    P. Eichler carried out this work as a member of the Saarbr\"ucken Graduate School of Computer Science.
    This research was funded in whole or in part by the German Research Foundation (DFG) grant 497132954.

    \subsubsection{\discintname}
    The authors have no competing interests to declare that are relevant to the content of this article.
\end{credits}

\bibliographystyle{splncs04}
\bibliography{references}

\newpage
\appendix
\crefalias{section}{appendix}
\crefalias{subsection}{subappendix}
\crefalias{subsubsection}{subsubappendix}
\section{Conjunctive Guards and Blocking Receives}\label{sec:blocking-guarded}

Our blocking semantics and conjunctive guards are closely related.
Specifically, the blocking semantics can be used to simulate the effects of a conjunctive guard:
For a concrete action $\act$ that is guarded with set of locations $\Locs_\forall$, we can remove all local actions of $\act$ from locations that do not appear in the guard $\Locs \setminus \Locs_\forall$.
As a consequence, $\act$ can only be enabled iff all processes are in $\Locs_\forall$.

If an action appears in combination with different conjunctive guards, we can copy the action with all of its local actions, assign each copy a unique name, and perform the removal for the different set of locations on the copy.

\begin{proposition}
    Synchronous actions with partially defined receive actions under the blocking semantics can simulate conjunctive guards.
\end{proposition}

Similarly, conjunctive guards can be used to simulate a blocking semantics by adding guards that ensure all processes are in locations where receives are defined.
We instead chose the blocking semantics, as it simplifies the presentation in the paper and the specifications.

\section{Appendix: Recovering Decidability}

This section provides more details on the details of the bounded model checking procedure presented in \cref{sec:bmc}.

\subsection{Complete SMT Encoding}\label{sec:complete-smt}

We start by completing the SMT formula $\phiacc$ which ensures that on every synchronous step the conditions three, four, and seven of \cref{sec:semantics}  are satisfied.

The formula is again split into three parts.
The first part $\philoc$, which constrains the acceleration factor choices for outgoing local actions to match the number of processes present in a location $\loc$, \ie it implements the constraint three in \cref{sec:semantics}:

\begin{align*}
    \philoc(\config, \Accfacs!, \Accfacs?, \loc, \act) := &\,\,\, \config.\vk[\loc] \! = \sum\limits_{(\loc,\act_i!!,\loc') \in \act.\send}\! \accfac_{\act_i!!} + \sum\limits_{\localfac=(\loc,\act??,\loc') \in \act.\recv}\! \accfac_{\localfac}
\end{align*}

$\phisuc$ then constrains the successor configuration $\config'$, determining how many processes are in location $\loc$ after $\act$ is executed,  encoding constraint four of the semantics of synchronous actions: 

\begin{align*}
    \phisuc(\config', \Accfacs!, \Accfacs?, \loc, \act)
    := &\,\,\, \config'.\vk[\loc]\! = \sum\limits_{(\loc',\act_i!!,\loc) \in \act.\send}\! \accfac_{\act_i!!} + \sum\limits_{\localfac=(\loc',\act??,\loc) \in \act.\recv}\! \accfac_{\localfac}
\end{align*}

$\phiacc$ then requires that $\philoc$ and $\phiacc$ hold for all locations and also ensures that the upper and lower bounds on the number of senders associated with action $\act$ are satisfied (\ie constraint 4 in \cref{sec:semantics}):

\begin{align*}
    \phiacc(\config, \config', \Accfacs!, \Accfacs?, \act) := & \sum \Accfacs! \geq \act.\low  \wedge \sum \Accfacs! \leq \act.\hi                                                                                                    \\
                                                             & \, \wedge \bigwedge\limits_{\loc \in \Locs} \big( \philoc(\config, \Accfacs!, \Accfacs?, \loc, \act) \wedge \phisuc(\config', \Accfacs!, \Accfacs?, \loc, \act) \big)
\end{align*}

\subsection{Context Switches in \TAS}\label{sec:context-changes-sync-ta}

Before providing a proof of the correctness of the constructed formula, we first examine effect of synchronous actions on the number of context switches on local rules.

As mentioned in \cref{sec:recap-bala}, for threshold automata, the encoding described in \cite{Balasubramanian20} relies on the fact that a threshold automaton with $k$ distinct guards can have at most $k + 1$ different contexts.
Here, a context refers to a set of guards being enabled or disabled along the execution.
As the guards are monotonic, as explained in \cref{sec:recap-bala}, for every guard this can only happen once during any run. 

Since synchronous steps preserve the monotonicity property of the shared variable guards and \cite[Proposition 1]{Balasubramanian20}, there can again be only $k$ context switches:

\begin{proposition}\label{prop:path-shape}
    Every path of a threshold automaton with synchronous steps with $k$ guards can have at most $k_\ta = k + 1$ context switches.
\end{proposition}

The design of $\phireach$ then allows up to $k_\ta$ context switches between every synchronous step.
This is because, in general, with only a bound on the number of synchronous steps, one cannot deduce constraints on the interleaving of synchronous steps and context switches.
\cref{ex:async-after-sync-step} demonstrates this.
In the future, we plan to investigate structural conditions to improve the size of the encoding.

\begin{example}\label{ex:async-after-sync-step}

    \begin{figure}[!ht]
        \centering
        \begin{subfigure}[t]{0.49\textwidth}
            \centering
            \begin{tikzpicture}[lts]
                \node[location,init] (q0) at (0,0) {$\loc_0$};

                \node[location] (q1) at (2,0) {$\loc_1$};

                \node[location] (q2) at (4,0) {$\loc_2$};

                \path[->]
                (q0) edge [loop above] node [] {$\var\varinc$} (q0)
                (q1) edge [loop above] (q1)
                (q0) edge [] node [] {$\act!!;\act??$} (q1)
                (q1) edge [] node [] {$x > 42$} (q2)
                ;
            \end{tikzpicture}

            \caption{\TAS where context switch needs to occur before synchronous step}
            \label{fig:ex-async-before-sync-step}
        \end{subfigure}
        \hfill
        \begin{subfigure}[t]{0.49\textwidth}
            \centering
            \begin{tikzpicture}[lts]
                \node[location,init] (q0) at (0,0) {$\loc_0$};

                \node[location] (q1) at (2,0) {$\loc_1$};

                \node[location] (q2) at (4,0) {$\loc_2$};

                \path[->]
                (q0) edge [loop above] (q0)
                (q1) edge [loop above] node [] {$\var\varinc$} (q1)
                (q0) edge [] node [] {$\act!!;\act??$} (q1)
                (q1) edge [] node [] {$x > 42$} (q2)
                ;
            \end{tikzpicture}

            \caption{\TAS where context switch needs to occur after synchronous step}
            \label{fig:ex-async-after-sync-step}
        \end{subfigure}

        \caption{\TAS with $k_\sync = 1$ and $k_\ta = 2$ but in \cref{fig:ex-async-before-sync-step} and \cref{fig:ex-async-after-sync-step} the context switches happen at different points of an execution reaching $\loc_2$.}
        \label{fig:ex-sync-async-step-timeline}
    \end{figure}

    Consider the \TAS depicted in \cref{fig:ex-sync-async-step-timeline}.
    For both TAs we can see that $k_\sync = 1$ (since there is only one synchronous transition and the transition is not part of a loop) and $k_\ta = 2$.

    In the first \TAS, on a run covering $\loc_2$, first the variable $\var$ must be increased using the self-loop on $\loc_0$ (which is an asynchronous transition) to enable the guard from $\loc_1$ to $\loc_2$.
    Only afterwards the synchronous transition from $\loc_0$ to $\loc_1$ should be taken, as incrementing $\var$ is not possible in $\loc_1$. 
    In this case the context switch must occur before executing a synchronous step.

    Conversely, for the TA depicted in \cref{fig:ex-async-after-sync-step}, the context switch can only appear after the synchronous step has been executed, as $\var$ can now only be increased in $\loc_1$.
\end{example}

\subsection{Correctness of the SMT Encoding}\label{sec:correct-smt}

We can now state the full correctness  proof for $\phireach$ presented in \cref{sec:bmc}.
We split the proof in three lemmas, the first two for the correctness of $\phisend$ and $\phimax$, the third one for formula $\phisync$ combining the two.
We then present the main correctness theorem for $\phireach$.

\begin{lemma}\label{thm:cor-send}
    Let $\syncta$ be a TA with synchronous actions, $\env$ its associated environment, $\CS{\env, \syncta}$ its counter system,
    $\act \in \globalActs$ an \tsender action and
    let $\config, \config'$ be two configurations of $\CS{\env, \syncta}$ with $\config.\vg = \config'.\vg$ and $\config.\vp = \config'.\vp$.
    Then $\config \trans{\act} \config'$ if and only if $\phisend(\config, \config',\act)$ evaluates to true.
\end{lemma}
\begin{proof}
    $\Rightarrow$:
    Let $\config \trans{\act} \config'$ be a valid synchronous step.
    Let $\Accfacs!, \Accfacs?$ be the acceleration factors chosen for this transition.

    We show that $\phisend$ is true because $\Accfacs!, \Accfacs?$ is a satisfying assignment for $\phisend$.
    $\bigwedge\limits_{\accfac \in \Accfacs!} n \geq 1$ must be satisfied because, by definition, at least one sender is required to execute every action.
    Left to show is that $\phiacc$ is satisfied:
    The first two conjuncts must be satisfied because the upper and lower guards of the action must be satisfied.
    The conditions $\philoc$ and $\phisuc$ must be satisfied by definition of the semantics \cref{sec:semantics} .

    $\Leftarrow$:
    Let $\Accfacs!,\Accfacs?$ be a valid assignment of $\phisend(\config, \config',\act)$.
    We show that $\Accfacs!,\Accfacs?$ is a valid set of acceleration factors by \cref{sec:semantics}.
    The first conjunction of $\phisend$ ensures that there is at least one sender per sending action.
    Then the other conditions of \cref{sec:semantics}are ensured by $\phiacc$:
    By definition, $\phisuc$ ensures the conditions on the successor configuration, $\philoc$ on the other hand ensures that the acceleration factors match the constraints of $\config$.
    The remaining conjuncts ensure that the send acceleration factors $\Accfacs!$ fulfill the upper and lower guards of the action.
    \qed
\end{proof}

\begin{lemma}\label{thm:cor-max}
    Let $\syncta$ be a threshold automaton with synchronous actions, $\env$ its associated environment, and let $\CS{\env,  \syncta}$ denote its counter system.
    Let $\config, \config'$ be two configurations of $\CS{\env, \syncta}$ with $\config.\vg = \config'.\vg$ and $\config.\vp = \config'.\vp$. 
    Then $\config \trans{\act} \config'$ with $\act$ a \tmaximal action if and only if there exists a valuation for which $\phimax(\config, \config', \act)$ evaluates to true.
\end{lemma}
\begin{proof}[Sketch]
    Analogously to \cref{thm:cor-send} except that the second conjunct ensures the condition $\min{\act.\hi, \!\!\sum\limits_{(\loc,a_i!!,\loc') \in \act.\send} \!\!\!\!\!\!\!\! \config.\vk[\loc]}$: 
    When the minimum is equal to $\act.\hi$ the first condition of the inner disjunction holds, if the minimum is determined by the number of available senders, only the second condition of the disjunction can become true, and it evaluates to true if and only if all available senders are used. 
    \qed
\end{proof}

\begin{lemma}\label{lem:cor-sync}
    Let $\syncta$ be a TA with synchronous actions, $\env$ its associated environment, and $\config, \config'$ be two configurations of $\CS{\env, \syncta}$ with $\config.\vp = \config'.\vp$.
    Then $\phisync(\config, \config')$ evaluates to true if and only if $\config = \config'$ or $\config \trans{\act} \config'$ for some $\act \in \globalActs$ of $\syncta$.
\end{lemma}
\begin{proof}[Sketch]
    Since synchronous actions cannot update shared variables, the conjunct $\config.\vg= \config'.\vg$ ensures that the variable valuations of $\config$ and $\config'$ are equal.
    Then $\phisync$ evaluates to true with $\sum S = 0$ iff $\config = \config'$.
    Otherwise, $\phisync$ evaluates to true if $\sum S = 1$ and therefore exactly one $s_\act = 1$.
    As $\phisend$ and respectively $\phimax$ by \cref{thm:cor-send} (\cref{thm:cor-max}) only evaluate to true if a valid synchronous step between $\config$ and $\config'$ exists $\phisync$ holds iff $\config \trans{\act} \config'$.
    \qed
\end{proof}

\begin{theorem}\label{thm:bmc-corr}
    Let $\syncta$ be a threshold automaton with synchronous actions and $k_\sync \in \Nat$.
    Then a configuration $\config'$ of $\syncta$ is reachable from configuration $\config$ with at most $k_\sync$ synchronous steps, if and only if $\phireach(k_\sync, \config, \config')$ is satisfiable.
\end{theorem}
\begin{proof}[Sketch]
    Let $\syncta$ be a threshold automaton with synchronous action and $k_\sync \in  \Nat$.
    Let $k_\ta = k + 1$ where $k$ denotes the number of distinct guards of local rules, see \cref{sec:context-changes-sync-ta}. 
    Every run from $\config$ to $\config'$ with $j \leq k_\sync$ synchronous steps can be divided along its synchronous steps:
    \[
        \config = \config_0 \trans{*} \config_0^{k_\ta} \trans{\act_0} \config_1 \trans{*} \dots \trans{\act_{j-1}} \config_j \trans{*} \config_j^{k_\ta} = \config',
    \]
    where every $\config_i \trans{*} \config_i^{k_\ta}$ is a (possibly empty) path only executing local rules.

    By correctness of $\phireachta$ \cite{Balasubramanian20}, $\phireachta(\config_i, \config_i^{k_\ta})$ holds iff such a path exists.
    \cref{lem:cor-sync} and \cref{prop:path-shape} implies that $\phisync(\config_i^{k_\ta}, \config_{i+1})$ holds iff $\config_i^{k_\ta} \trans{\act_i} \config_{i+1}$ or $\config_i^{k_\ta} = \config_{i+1}$.
    Since all paths can be divided into a form as stated above, it follows that $\phireach$ holds if and only if a path from $\config$ to $\config'$ with at most $k_\sync$ steps exists.
    \qed
\end{proof}

\subsection{Layering Algorithm}\label{sec:layering-alg}

After having established the correctness of our bounded model checking procedure, this section introduces a practical algorithm to compute layers in a \TAS.

\newcommand{\SCC}{\ensuremath{S}}
\begin{algorithm}
    \SetKwProg{Fn}{Function}{}{}
    \SetKwFunction{Weight}{weight}
    \SetKwFunction{TarjanSCC}{tarjanSCC}
    \SetKwFunction{isSink}{isSink}

    $\{ \SCC_0, \dots, \SCC_m \} \leftarrow \TarjanSCC{\syncta}$;
    $\sinkLayer = \emptyset$\;

    \tcp{Check $\syncta$ is layered and determine the sink layer}
    \For{$\SCC \in \SCC_0, \ldots, \SCC_m$}{
        \tcp{sync.\ transition between distinct locations of one SCC: cannot be layered}
        \If{$\exists \loc, \loc' \in \SCC. \, \loc \trans{\someact} \loc'$}{
            \tcp{Only receive self-loops are allowed}
            \eIf{$\loc = \loc' \wedge \isSink{\loc}$}{
                \tcp{Move to sink layer}
                 $\sinkLayer \leftarrow \sinkLayer \cup \{ \loc \}$\;
                 $\SCC \leftarrow S \setminus \{ \loc \}$\;
            }{
                \Return \texttt{ERROR}; \tcp{$\syncta$ is not layered}
            }
        }
    }

    \tcp{Process SCCs in topological order and assign them to layer}
    \For{$\SCC \in S_0, \ldots, \SCC_m$}{
        \eIf{$\SCC$ has no incoming transition}{
            $\layeridx(\SCC) \leftarrow 1$; \tcp{add to layer 1}
        }{
            \tcp{take max over index of incoming}
            $\layeridx(\SCC) \leftarrow \max{\{ \layeridx(\SCC')  + \Weight(\act) \mid \SCC' \neq \SCC \, \text{s.t.}\, \exists \loc \in \SCC', \loc' \in \SCC. \, \loc \trans{\someact} \loc' \}}$\;
        }
    }

    \tcp{Return layers as partition}
    $\Part \leftarrow []$\;
    \For{$i \in 1, \ldots, \max{ \{\layeridx(\SCC) \mid \SCC \in \SCC_0, \ldots, \SCC_m \wedge \SCC \neq \emptyset \}} + 1$}{
        $\layer_{i} \leftarrow \bigcup\limits_{\SCC \in S_0,\ldots, S_m \wedge \layeridx(\SCC) = i} \SCC$\;
        $\Part \leftarrow \Part \concat [\layer_i]$\;
    }
    \If{$\sinkLayer \neq \emptyset$}{
        $\Part \leftarrow \Part \concat [\sinkLayer]$\;
    }

    \Return $\Part$\;

    \vspace{5mm}

    \tcp{Weight of a rule: $0$ if asynchronous, $1$ if synchronous}
    \Fn{\Weight{$\act \in \localActs \cup \{ \noact \}$}}{
        \eIf{$\act = \noact$
            }{\Return 0
            }{\Return 1}
    }

    \vspace{5mm}

    \tcp{Compute SCCs of $\syncta$ using Tarjan's algorithm and return SCCs in topological order}
    \Fn{\TarjanSCC{$\syncta$}}{
        $\ldots$
    }

    \tcp{Check that $\loc$ is a sink location with only receive self-loops}
    \Fn{\isSink{$\loc$}}{
        $\ldots$
    }
    \caption{Compute layer partition of $\syncta = \langle \Locs, \initial{\Locs}, \SharedVars, \Rules, \globalRules \rangle$}
    \label{alg:partition}
\end{algorithm}

The algorithm in pseudo code is presented in \cref{alg:partition}. It works by first computing the strongly connected components (SCCs) of the underlying graph of the automaton and sorting them in topological ordering.
This can be achieved by utilizing Tarjan's algorithm \cite{Tarjan72} and reversing the order of SCCs computed.

In a layered \TAS, all locations of an SCC must be in the same layer, because every location $\loc$ of an SCC is reachable from every other location $\loc'$ of an SCC and vice versa.
Therefore, if one were to assign $\loc$ and $\loc'$ to two different layers and assign an order between those layers, there would be some path from a higher layer to a lower layer, which directly violates the layeredness assumption.

Note that since all locations inside an SCC are always part of the same layer, by \cref{def:layered}, within one SCC no synchronous transitions are allowed.

After explaining the intuition behind the correctness of \cref{alg:partition}, we can formally state the correctness argument in detail:

\lemLayeredScc*
\begin{proof}[Sketch]
    \newcommand{\weight}{\ensuremath{\mathtt{weight}}}
    $\Rightarrow$:
    Proof by contradiction.
    Let $\syncta$ be threshold automaton with synchronous steps $\syncta$, and $\Part$ be a valid partitioning of $\syncta$ according to \cref{def:layered}.

    Suppose there exists an SCC $S$ with a synchronous transition $\loc \trans{\act} \loc'$ with $\loc,\loc' \in S$, $\act \in \localActs$ which is not a receive self-loop on a sink location.

    By definition of a layered automaton (\cref{def:layered}), $\loc$ must be in a lower layer than $\loc'$ (as synchronous transition can only transition to a higher layer).

    However, since $\loc$ and $\loc'$ are part of the same SCC there must also exist a path from $\loc'$ to $\loc$.
    On this path, there must consequently exist a transition (asynchronous or synchronous) that transitions to a lower layer.
    Such a transition is not allowed within a valid partition by \cref{def:layered}, which contradicts our assumption.

    $\Leftarrow$:
    Let $\syncta$ threshold automaton for which the underlying graph SCC contains no synchronous transition within an SCC (except for receive self-loops in sink locations).
    We prove that the automaton is layered by constructing a valid partition using \cref{alg:partition}.

    Note that all locations, which only have receive self-loops and are sink locations, are moved to $\sinkLayer$ by the instructions in lines 4 to 6. 
    Therefore, next we have to consider the computation of the other layers.

    Let $S_0, \dots, S_m$ be the SCCs in topological order (which is ensured by Tarjan's algorithm \cite{Tarjan72}) of the graph of the automaton.

    \cref{alg:partition} assigns each SCC $S$ to the layer index as one for all source layers and 
    \[
        \layeridx(S) = \max{\{\, \layeridx(S') + \weight(\act)
        \mid \exists \loc \in S',\, \loc' \in S.\; \loc \trans{\act} \loc' \,\}},
    \]
    where $\weight(\noact) = 0$ and $\weight(\act) = 1$ for $\act \in
    \localActs$, otherwise. 

    Since the algorithm starts with the source SCCs as the SCCs are processed in topological order, \ie  any predecessor of an SCC must have been already processed before the SCC, as otherwise the SCCs would not be topologically ordered.
    
    Note that in particular in a topological ordering of SCCs, there are no edges to a previous SCC.

    Therefore, this processing order also ensures that the two conditions of \cref{def:layered} are upheld:
    \begin{itemize}
        \item \emph{Asynchronous rule.}
        For $\loc \trans{\noact} \loc'$ with $\loc \in S$, $\loc' \in S'$:
        $\layeridx(S') \geq \layeridx(S) + 0$, so $\loc'$ is in the same or a
        higher layer. 
        If this was not the case, there would exist an edge into a previous SCC and therefore topological ordering would be violated. 

        \item \emph{Synchronous rule.}
        For $\loc \trans{\act} \loc'$ with $\act \in \localActs$:
        by assumption $\loc$ and $\loc'$ lie in distinct SCCs, so
        $\layeridx(S') \geq \layeridx(S) + 1 > \layeridx(S)$, meaning $\loc'$
        is in a strictly higher layer.
        If this was not the case, there would exist either an edge into a previous SCC and therefore topological ordering would again be violated or there would be a synchronous transition within one SCC violating our assumption.
    \end{itemize}
    Hence $\syncta$ is layered.
\end{proof}

\section{Appendix: Modular Models \& Replacement}

\subsection{SMT Encoding Reachability-Preserving Replacements}\label{sec:smt-reach-preserve}

This section provides the SMT encodings omitted in \cref{sec:modular-replace}.

We start with the definition of $\phiinit(\config)$ which is true if and only if a configuration $\config$ is an initial configuration:
\begin{align*}
    \phiinit(\config) := & \sum\limits_{\loc \in \initial{\Locs}} \config.\vk[\loc] = \N(\config.\vp) \wedge \bigwedge\limits_{\loc \in \Locs \setminus \initial{\Locs}} \config.\vk[\loc] = 0 \wedge \bigwedge_{\var \in \SharedVars} \config.\vg[\var] = 0
\end{align*}

Next, the following theorem provides the encoding for checking whether a replacement is strongly reachability-preserving:
\begin{theorem}\label{thm:strong-reach}
    Let $\syncta$ and $\replace{\syncta}{\replacement}$ be layered \TAS with $k_\sync = |\Part_\syncta|{-}1$ and $k'_\sync=|\Part_{\replace{\syncta}{\replacement}}|{-}1$. Then $\replacement$ is strongly reachability-preserving iff  the following formula holds:
    \begin{align*}
        \phieqstrong(\syncta, \replacement) := & \, \phieq(\syncta,\replacement) \wedge \forall \initial{\config},\config \in \Configs_{\replace{\syncta}{\replacement}}. \,
        \big( \phiinit(\initial{\config}) \wedge
        \phireach(k'_\sync,\initial{\config}, \config)                                                                                                                       \\
                                              & \wedge \bigwedge\limits_{\LocsAdd}^\loc \config.\vk[\loc] = 0
        \implies \exists \initial{\config'},\config' \in \Configs_{\syncta}. \, \phiinit(\initial{\config'}) \wedge \phireach(k'_\sync,\initial{\config'},\config')          \\
                                              & \wedge
        \bigwedge\limits_{\Locs \setminus \LocsRM}^\loc \config.\vk[\loc]  = \config'.\vk[\loc] \wedge
        \bigwedge\limits_{\LocsRM}^\loc \config'.\vk[\loc] = 0
        \wedge  \!\bigwedge\limits_{\SharedVars \setminus \VarsRM}^{\var} \!\! \config.\vg[\var] = \config'.\vg[\var]                                               \\
                                              & \wedge \config.\vp = \config'.\vp
        \big)
    \end{align*}
\end{theorem}

We can now also describe more precisely which properties transfer after a reachability-preserving replacement:

\begin{lemma}\label{lem:replacement}
    Let $\syncta$ be a threshold automaton with synchronous actions and $\replacement$ a reachability-preserving replacement, $\env$ the environment, and let $\varphi$ be a cardinality constraint, which only references locations in $\Locs_\syncta \setminus \LocsRM$.
    If a configuration satisfying $\varphi \wedge \bigwedge_{\loc \in \LocsRM} \#\loc = 0$ is reachable in $\syncta$,
    then a configuration satisfying $\varphi \wedge \bigwedge_{\loc \in \LocsAdd} \#\loc = 0$ is reachable in
    $\replace{\syncta}{\replacement}$.
    If $\replacement$ is strongly reachability preserving, the converse direction holds as well.
\end{lemma}
\begin{proof}[Sketch]
    Let $\config$ be a reachable configuration of $\syncta$ with $\config \models \varphi \wedge \bigwedge_{\loc \in \LocsRM} \#\loc = 0$.
    By \cref{def:reach-preserve}, there is a reachable configuration $\config'$ of $\replace{\syncta}{\replacement}$ with
    $\config'.\vk[\loc] = 0$ for all $\loc \in \LocsAdd$ and $\config'.\vk[\loc] = \config.\vk[\loc]$ for all $\loc \in \Locs \setminus \LocsRM$.
    As $\varphi$ only constrains locations in $\Locs \setminus \LocsRM$, also $\config' \models \varphi$.
    For strong reachability-preserving replacements, the other direction follows analogously.
    \qed
\end{proof}

\subsection{Local Replacement Condition}\label{sec:local-replace-cond}

We have seen, that the SMT formulas for checking whether a replacement is (strong) reachability preserving essentially encode the whole reachability relation of both automata.
This is necessary because the replacement definition is very general, in practice replacements only concern a limited set of the locations of an automaton.

To obtain a more local property, we examine the benchmarks and make the following observations:
\begin{itemize}
    \item Modules are typically replaced at the beginning of an automaton.
          In our examples, the algorithms are usually built upon the results of the execution of a previous module.
          Consequently, the replacement usually starts at the initial states, \ie $\initial{\Locs} = \LocsIn$.

    \item Processes pass through the module at most once.
          Most algorithms do not need to execute a module multiple times, after the processes leave the state space of the module they cannot return.
\end{itemize}

We first define $\replaceTA{\syncta}{\replacement}$ which is exactly the (sub-) automaton which gets replaced by a replacement $\replacement$.

\newcommand{\initRLoc}{\ensuremath{\Locs_{\mathsf{init},\replacement}}}
\begin{definition}
    For a \TAS $\syncta = \langle \Locs, \initial{\Locs}, \SharedVars, \Rules, \globalRules \rangle$ and a replacement $\replacement = \langle \LocsRM, \allowbreak\LocsAdd, \RulesAdd, \globalRulesAdd, \VarsRM \rangle$.
    Let $\Rules_\replacement = \{ \rl \in \Rules \mid \rl.\vfrom \in \LocsRM  \vee \rl.\vto \in \LocsRM \}$,
    $\globalRules_\replacement = \{ (\act,\_,\_,\_,\_,\_) \in \globalRules \mid \exists (\loc,\_, \loc') \in \act.\send \cup \act.\recv. \, \loc \in \LocsRM  \vee \loc' \in \LocsRM \}$ and $\Locs_\replacement = \LocsRM \cup \LocsIn \cup \LocsOut$.
    We denote by $\replaceTA{\syncta}{\replacement} = \langle \Locs_\replacement, \initial{\Locs}, \SharedVars, \Rules_\replacement, \globalRules_\replacement \rangle$ the \emph{replacement automaton}.
\end{definition}

\newcommand{\vars}{\ensuremath{\mathit{vars}}}
\newcommand{\acts}{\ensuremath{\mathit{acts}}}
\newcommand{\tgt}{\ensuremath{\mathit{tgt}}}
\newcommand{\globalRulesPost}{\ensuremath{\globalRules_{\mathsf{post}}}}
\newcommand{\RulesPost}{\ensuremath{\Rules_{\mathsf{post}}}}

We also define
\begin{itemize}
    \item $\RulesPost = \Rules \cap \Rules'$, where $\Rules'$ denotes the rules of $\replace{\syncta}{\replacement}$, \ie the rules which are not updated by the replacement, and
    \item $\globalRulesPost = \globalRules \cap \globalRules'$, where $\globalRules'$ denotes the rules of $\replace{\syncta}{\replacement}$, \ie the rules which are not updated by the replacement.
\end{itemize}
Also, let $\vars(\Rules)$ denote the shared variables appearing on $\Rules$, $\acts(\globalRules)$ the set of global actions appearing on $\globalRules$, and, for a set $T$ of rules and global rules, let $\mathit{src}(T)$ and $\tgt(T)$ denote the sets of all source and target locations of the rules and local transitions in $T$.

\begin{definition}\label{def:local-reach}
    Let $\syncta$ be an automaton, $\replacement$ a replacement, we call $\replacement$ \emph{local reach-ability-preserving} if it satisfies (R1)-(R4) below and $\replacement$ is reachability-preserving on $\replaceTA{\syncta}{\replacement}$.
    If $\replacement$ satisfies R1-R4 below and is strongly reachability-preserving on $\replaceTA{\syncta}{\replacement}$, we call it \emph{strongly local reachability-preserving}.
    \begin{itemize}
        \item(R1) $\initial{\Locs} = \LocsIn$, \ie the replaced part is the beginning of the automaton.
        \item(R2) $\tgt(\RulesPost \cup \globalRulesPost) \cap \LocsIn = \emptyset$,
              $\mathit{src}(\Rules_\replacement \cup \RulesAdd \cup \globalRules_\replacement \cup \globalRulesAdd) \subseteq \LocsIn \cup \LocsRM \cup \LocsAdd$, and
              $\tgt(\Rules_\replacement \cup \RulesAdd \cup \globalRules_\replacement \cup \globalRulesAdd) \subseteq \Locs_\replacement \cup \LocsAdd$,
              \ie the replaced part only acts inside the module, and once a process has left $\LocsIn \cup \LocsRM \cup \LocsAdd$, it never returns.
        \item(R3) $\vars(\Rules_\replacement \cup \RulesAdd) \cap \vars(\RulesPost) = \emptyset$,
              \ie the replaced part does not share variables with the rest of the automaton.
        \item(R4) $\acts(\globalRules_\replacement \cup \globalRulesAdd) \cap \acts(\globalRulesPost) = \emptyset$,
              \ie the replaced part does not share synchronous actions with the rest of the automaton.
    \end{itemize}
\end{definition}

R1 - R4 are simple syntactic checks on the automaton, and the reachability-preserving conditions now only have to be checked on the parts of the automaton that are replaced.
Depending on the size of the replacement compared to the original automaton, this can drastically reduce the size of the encoding.

Next we show that checking local reachability preserving is enough:

\begin{lemma}\label{lem:local-reach-preserving}
    Let $\env$ be an environment, $\syncta$ be an automaton and  $\replacement$ a replacement in $\syncta$.
    If $\replacement$ is (strong) local reachability-preserving then $\replacement$ is also (strong) reachability preserving.
\end{lemma}
\begin{proof}
    Proof by contradiction,
    \ie assume there exists a reachable configuration $\config$ (with $\LocsRM$ empty) in $\syncta$ such that no matching configuration is reachable in $\replace{\syncta}{\replacement}$.
    Let $\run = \config_0 \trans{} \ldots \trans{} \config$ denote the computation reaching $\config$.

    We show that it is possible to find a run $\run'$ in $\replace{\syncta}{\replacement}$ reaching a configuration equivalent to $\config$, contradicting the assumption:

    We first show that we can reorder and split the run $\run$ into two parts $\run_1$, $\run_2$ where $\run_1$ only executes rules that are replaced by $\replacement$ and the second half $\run_2$, only executes rules that are not replaced.

    \smartpar{Reordering the run}
    By R2 and the definition of $\RulesPost$ and $\globalRulesPost$, rules that are not replaced never move a process into $\LocsIn \cup \LocsRM$, and all replaced rules start in $\LocsIn \cup \LocsRM$.
    Note that as soon as a process moves out of $\LocsIn \cup \LocsRM$ it disables all actions of $\globalRules_\replacement$ for the rest of the run.

    Assume the run $\run$ is not yet in the shape $\run = \run_1 \run_2$, \ie there must exist a step which executes a rule in $\RulesPost$ or $\globalRulesPost$, which is followed by a step of a rule or action removed by the replacement.

    The step using a replaced rule cannot be synchronous, since for the previous step to have been enabled, there must have been some process in $\Locs \setminus \LocsRM$, which stays outside the module by R2. 
    By R3 it follows that every synchronous action must be disabled, by the process, since the process has no corresponding receive action.
    Similarly, the first step executing  a rule  from $\RulesPost$ and $\globalRulesPost$ cannot be synchronous either, since it would require all processes to be in $\Locs \setminus \LocsRM$ (or there would again be some blocking process), disabling any step using a replaced rule. 
    Hence, both steps are asynchronous.
    The source of the second step lies in $\LocsIn \cup \LocsRM$ and is therefore not the target of the first step, and by R4 neither step can update a variable of the other's guard, so we can swap the execution of the rules.
    Let $\config_{\mathsf{mid}}$ be the configuration between $\run_1$ and $\run_2$.
    As $\run_2$ does not move processes into or out of $\LocsRM$, also $\config_{\mathsf{mid}}.\vk[\loc] = 0$ for all $\loc \in \LocsRM$.

    \smartpar{Simulating $\run_1$}
    By R1 $\config_0$ is an initial configuration of $\replaceTA{\syncta}{\replacement}$ and $\replace{\syncta}{\replacement}$, and by R2 $\run_1$ is a run of $\replaceTA{\syncta}{\replacement}$.
    Since the replacement is local reachability preserving, we therefore know that there must exist a run reaching a configuration equivalent to $\config_{\mathsf{mid}}$ in $\replace{\replaceTA{\syncta}{\replacement}}{\replacement}$ and since by R2 all processes remain in the locations of $\replace{\replaceTA{\syncta}{\replacement}}{\replacement}$, the run consequently also has to exist in $\replace{\syncta}{\replacement}$.
    We refer to this run as $\run_1'$.

    \smartpar{Simulating $\run_2$}
    Since $\run_1'$ ends in a configuration equivalent to $\config_{\mathsf{mid}}$, we can replay $\run_2$ step by step.
    By definition, rules in $\RulesPost$ and $\globalRulesPost$ do not interfere with $\LocsRM \cup \LocsAdd$.
    Therefore, they remain in $\replace{\syncta}{\replacement}$ and only refer to locations and variables on which both agree.
    By R4, both automata have the same rules for every action of $\globalRulesPost$, so synchronous steps remain enabled as well.
    Hence, the resulting run reaches a configuration equivalent to $\config$, contradicting our assumption.

    The strong direction follows analogously.
    \qed
\end{proof}

\section{Appendix: Evaluation \& Case Study}

\subsection{TA Models \& Synchronous Abstractions}\label{sec:missing-models}

This section will provide more detail on the king consensus and king consensus algorithms, and it will introduce the remaining threshold automata and synchronous models, and the properties that have been checked for each.
Again the algorithm representations and correctness properties are taken from \cite{FitziM00}.

\smartpar{Weak Consensus}
The weak consensus TA has been discussed in \cref{ex:weak-cons-ta-intro} and its synchronous abstraction in \cref{ex:weak-cons-sync}.
The correctness properties of weak consensus are
\begin{itemize}
    \item \textbf{Weak Consistency}:  If the output $y_i \in  \{0,1\}$ for some honest player $\player_i$, then $y_j \in \{ \bot, y_i \}$ for all other honest players $\player_j$. We translate this to $\globally(y_0 > 0 \Rightarrow y_1 = 0)$ and $\globally(y_1 > 0 \Rightarrow y_0 = 0)$.
    \item \textbf{Weak Persistency}: If all honest players have the same input value $x$, then all honest players output $y_i = x$. We translate this to $x_1 = 0 \Rightarrow \globally(y_1 = 0 \wedge y_\bot = 0)$ and $x_0 = 0 \Rightarrow \globally(y_0 = 0 \wedge y_\bot = 0)$.
\end{itemize}

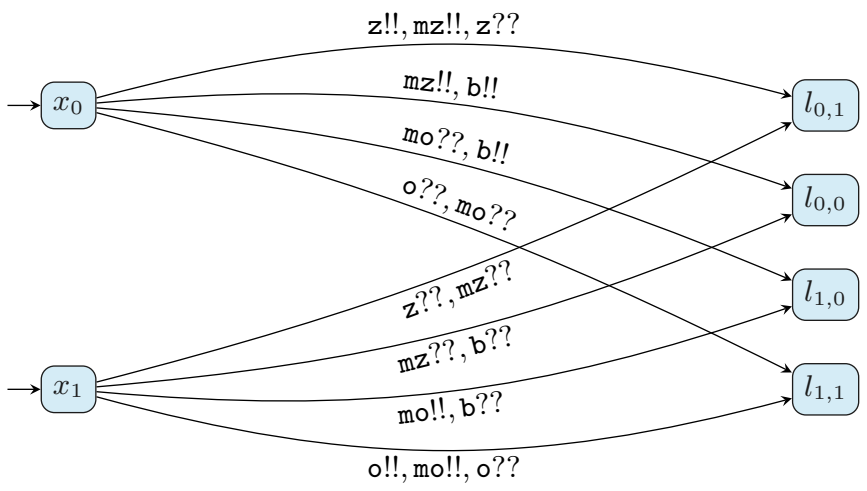
\begin{figure}
    \centering
    \resizebox{0.49\textwidth}{!}{
        \begin{tikzpicture}[lts]
    \node (i0) [location,init] at (0, 0) {$x_0$};
    \node (i1) [location,init] at (0, -3){$x_1$};
    \node (01) [location] at (8, 0) {$\loc_{0,1}$};
    \node (00) [location] at (8, -1) {$\loc_{0,0}$};

    \node (10) [location] at (8, -2) {$\loc_{1,0}$};
    \node (11) [location] at (8, -3) {$\loc_{1,1}$};

    \path[->]
        (i0) edge [bend left=15] node [above, sloped, yshift=-2pt] {$\actdec{z}!!,\actdec{mz}!!,\actdec{z}??$} (01)
        (i1) edge [ bend right=15] node [below, sloped, yshift=1.5pt] {$\actdec{o}!!,\actdec{mo}!!,\actdec{o}??$} (11)

        (i0) edge [bend left=12] node [above, sloped, yshift=-2pt] {$\actdec{mz}!!,\actdec{b}!!$} (00)
        (i1) edge [bend right=12] node [below, sloped, yshift=1.5pt] {$\actdec{mo}!!,\actdec{b}??$} (10)

        (i0) edge [bend left=9] node [above, sloped, yshift=-2pt] {$\actdec{mo}??,\actdec{b}!!$} (10)
        (i1) edge [bend right=9] node [below, sloped, yshift=1.5pt] {$\actdec{mz}??,\actdec{b}??$} (00)

        (i0) edge [bend left=6] node [above, sloped, yshift=-2pt] {$\actdec{o}??,\actdec{mo}??$} (11)
        (i1) edge [bend right=6] node [below, sloped, yshift=1.5pt] {$\actdec{z}??,\actdec{mz}??$} (01)
    ;
\end{tikzpicture}
    }
    \caption{Synchronous overapproximation of graded consensus with $\actdec{z},\actdec{o}$ \tsender actions with lower bound $n{-}t{-}f$ and $\actdec{mz}, \actdec{mo}$ \tmaximal actions with lower bounds $n{-}t{-}f$, upper bound $n{-}t{-}1$ and $\actdec{b}$ also a \tmaximal action with lower bound $t{-}f{+}1$ and upper bound $n{-}t{-}1$.}
    \label{fig:graded-sync-over}
\end{figure}

\smartpar{Graded Consensus}
We start with the synchronous overapproximation of the graded consensus algorithm presented in \cref{fig:graded-sync-over}.
Similar to the weak consensus algorithm introduced in \cref{ex:weak-cons-sync}, the actions $\actdec{z}$ and $\actdec{o}$ represent the acceptance of value zero or one, and $\actdec{mz}, \actdec{mo}$ the case where one value might be accepted by some processes using the votes of Byzantine processes.
Again, the action $\actdec{b}$ can fire when no value can be accepted with grade one.
In this case, $\actdec{b}$ moves processes directly to $\loc_{0,0}$ or $\loc_{1,0}$, so processes can reach $\loc_{1,0}$ even if $f=0$.

\begin{figure}
    \centering
    \resizebox{0.5\textwidth}{!}{
        \begin{tikzpicture}[lts]
    \node (i0) [location,init] at (0, 0) {$x_0$};
    \node (i1) [location,init] at (0, -4.5){$x_1$};

    \node (01) [location] at (10, 0) {$\loc_{0,1}$};
    \node (00) [location] at (10, -1.25) {$\loc_{0,0}$};
    \node (b) [location] at (7.5, -2.25) {$\loc_{\bot}$};
    \node (10) [location] at (10, -3.25) {$\loc_{1,0}$};
    \node (11) [location] at (10, -4.5) {$\loc_{1,1}$};

    \path[->]
        (i0) edge [bend left=18] node [above, sloped, yshift=-2pt] {$\actdec{z}!!,\actdec{mz}!!,\actdec{z}??$} (01)
        (i1) edge [bend right=18] node [below, sloped, yshift=1.5pt] {$\actdec{o}!!,\actdec{mo}!!,\actdec{o}??$} (11)

        (i0) edge [bend left=15] node [above, sloped, yshift=-2pt] {$\actdec{mz}!!$} (00)
        (i1) edge [bend right=15] node [below, sloped, yshift=1.5pt] {$\actdec{mo}!!$} (10)

        (i0) edge [bend left=9] node [above, sloped, yshift=-2pt] {$\actdec{b}!!$} (b)
        (i1) edge [bend right=9] node [below, sloped, yshift=1.5pt] {$\actdec{b}??$} (b)

        (i0) edge [bend left=21] node [above, sloped, yshift=-2pt] {$\actdec{mo}??$} (10)
        (i1) edge [bend right=21] node [below, sloped, yshift=1.5pt] {$\actdec{mz}??$} (00)

        (i0) edge [bend left=6] node [above, sloped, yshift=-2pt] {$\actdec{o}??,\actdec{mo}??$} (11)
        (i1) edge [bend right=6] node [below, sloped, yshift=1.5pt] {$\actdec{z}??,\actdec{mz}??$} (01)

        (b) edge [] node [above, sloped, yshift=-2pt] {$\actdec{k}!!,\actdec{k}??,\actdec{fl}??$} (00)
        (b) edge [] node [below, sloped, yshift=1.5pt] {$\actdec{fl}!!,\actdec{fl}??$} (10)
    ;
\end{tikzpicture}
    }
    \caption{Exact synchronous model of graded consensus with $\actdec{z},\actdec{o},\actdec{mz},\actdec{mo}$ as in \cref{fig:graded-sync-over}, $\actdec{b}$ a \tmaximal action with $\actdec{b}.\low = t{-}f{+}1$ and $\actdec{b}.\hi = n{-}t{-}1$, and $\actdec{k}, \actdec{fl}$ \tsender actions with $\actdec{k}.\low = \actdec{fl}.\low = 1$ and $\actdec{fl}.\hi = f$.}
    \label{fig:graded-sync-exact}
\end{figure}

\cref{fig:graded-sync-exact} shows an exact synchronous model, which coincides with the overapproximation except for the action $\actdec{b}$.
In this case, $\actdec{b}$ moves all processes to the intermediate location $\loc_\bot$.
From there, either the action $\actdec{k}$ moves all processes to $\loc_{0,0}$, or the action $\actdec{fl}$ moves all its senders to $\loc_{1,0}$ and all other processes to $\loc_{0,0}$ or $\loc_{1,0}$.
However, since action $\actdec{fl}$ has upper bound $f$, it can only be enabled if $f \geq 1$ and therefore the model is exact.

The graded consensus guarantees the following properties:
\begin{itemize}
    \item \textbf{Graded Consistency}: If an honest player $\player_i$ outputs $(y_i, 1)$, then $y_j = y_i$ for all other honest players $\player_j$. We translate this to $\globally(\loc_{0,1} > 0 \Rightarrow \loc_{1,0} = 0 \wedge \loc_{1,1} = 0)$ and $\globally(\loc_{1,1} > 0 \Rightarrow \loc_{0,0} = 0 \wedge \loc_{0,1} = 0)$.
    \item \textbf{Graded Persistency}: If all honest players have the same input value $x$, then all honest players output $(y_i, g_i) = (x, 1)$. We translate this to $x_1 = 0 \Rightarrow \globally(\loc_{0,0} = 0 \wedge \loc_{1,0} = 0 \wedge \loc_{1,1} = 0)$ and $x_0 = 0 \Rightarrow \globally(\loc_{0,0} = 0 \wedge \loc_{0,1} = 0 \wedge \loc_{1,0} = 0)$.
\end{itemize}

\begin{figure}
    \centering
    \begin{minipage}[c][50mm]{0.45\textwidth}
        \centering
        \input{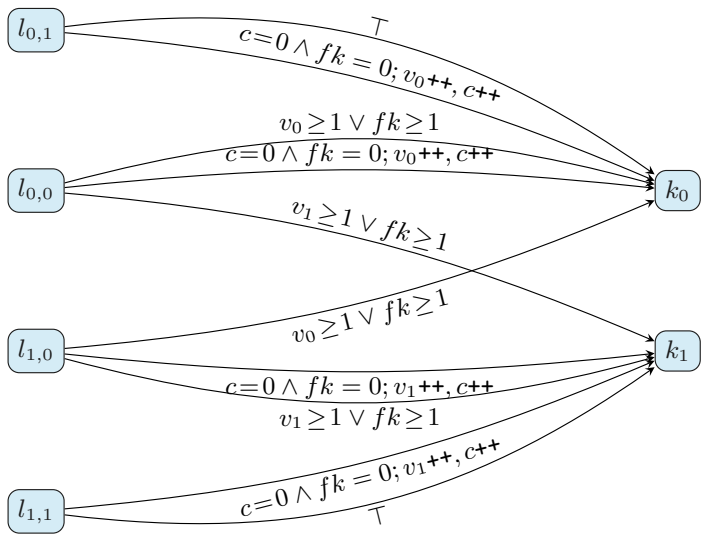}
    \end{minipage}
    \hfill
    \begin{minipage}[c][50mm]{0.39\textwidth}
        \centering
        \resizebox{\textwidth}{!}{
            \begin{tikzpicture}[lts]
    \node (01) [location] at (0, 0) {$\loc_{0,1}$};
    \node (00) [location] at (0, -2) {$\loc_{0,0}$};

    \node (10) [location] at (0, -4) {$\loc_{1,0}$};
    \node (11) [location] at (0, -6) {$\loc_{1,1}$};

    \node (o0) [location] at (8,-2) {$k_0$};
    \node (o1) [location] at (8,-4) {$k_1$};

    \path[->]
        (01) edge [bend left=21] node [above, sloped, yshift=-2pt] {$\top$} (o0)
        (01) edge [bend left=9] node [above, sloped, yshift=-2pt] {$c \!=\! 0 \wedge fk = 0; v_0\varinc , c\varinc$} (o0)
        (11) edge [bend right=21] node [below, sloped, yshift=1.5pt] {$\top$} (o1)
        (11) edge [bend right=9] node [below, sloped, yshift=1.5pt] {$c \!=\! 0 \wedge fk = 0; v_1\varinc , c\varinc$} (o1)
        
        (00) edge [bend left=15] node [above, sloped, yshift=-2pt] {$v_0 \!\geq\! 1 \vee fk \!\geq\! 1$} (o0)
        (10) edge [bend right=15] node [below, sloped, yshift=1.5pt] {$v_1 \!\geq\! 1 \vee fk \!\geq\! 1$} (o1)
 
        (00) edge [bend left=6] node [above, sloped, yshift=-2pt] {$c \!=\! 0 \wedge fk = 0; v_0\varinc , c\varinc$} (o0)
        (10) edge [bend right=6] node [below, sloped, yshift=1.5pt] {$c \!=\! 0 \wedge fk = 0; v_1\varinc , c\varinc$} (o1)

        (00) edge [bend left=9] node [above, sloped, yshift=-2pt] {$v_1 \!\geq\! 1 \vee fk \!\geq\! 1$} (o1)
        (10) edge [bend right=9] node [below, sloped, yshift=1.5pt] {$v_0 \!\geq\! 1 \vee fk \!\geq\! 1$} (o0)
    ;

\end{tikzpicture}
        }
        \caption{King consensus threshold automaton $A_{K}$, where $\Params = \{ n, t, f, pk \}$ and $\RC = n > 3t \wedge t \geq f \wedge pk \leq 1$ and initial condition $pk = vk$.}
        \label{fig:king-cons-ta}
    \end{minipage}
\end{figure}

\newpage
\smartpar{King Consensus}

\begin{wrapfigure}[13]{r}{0.425\linewidth}
    \vspace{-12mm}
    \centering
    \resizebox{0.9\linewidth}{!}{
        \begin{tikzpicture}[lts]
    \node [location, init] (i0) at (0,0) {$x_0$};
    \node [location, init] (i1) at (0,-1.5) {$x_1$};
    
    \node [location] (o0) at (4,0) {$k_0$};
    \node [location] (o1) at (4,-1.5) {$k_1$};

    \path[->]
        (i0) edge [bend left=15] node [above, sloped, yshift=-2pt] {$\actdec{z}!!,\actdec{b}!!,\actdec{z}??,\actdec{b}??$} (o0)
        (i1) edge [bend right=15] node [below, sloped, yshift=1.5pt] {$\actdec{o}!!,\actdec{b}!!,\actdec{o}??,\actdec{b}??$} (o1)

        (i0) edge [bend left=12] node [above,sloped, yshift=-2pt] {$\actdec{o}??,\actdec{b}??$} (o1)
        (i1) edge [bend right=12] node [below, sloped, yshift=1.5pt] {$\actdec{z}??,\actdec{b}??$} (o0)
    ;

\end{tikzpicture}
    }
    \captionsetup{justification=raggedright}
    \caption{King consensus synchronous module with $\actdec{z}$, $\actdec{o}$ \tsender actions with lower bound $1$ and $\actdec{b}$ also an \tsender action with lower bound $2$ and upper bound $2 * pk$}
    \label{fig:king-sync}
\end{wrapfigure}
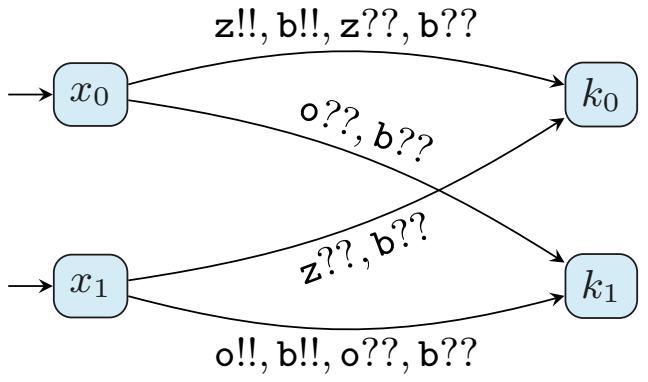

\cref{fig:king-cons-ta} depicts the additional locations that a threshold automaton adds on top of the graded consensus automaton.
Note that the model uses an additional parameter $pk$, with an additional initial condition $pk = vk$.
The parameter is used to model whether the algorithm assumes an honest king (if $pk = 0$ and therefore $vk = 0$) or whether the round assumes a faulty king ($pk = vk = 1$), which can send different king votes to every process.

The synchronous abstraction of a king consensus algorithm is depicted in \cref{fig:king-sync} and only has three actions, $\actdec{o}$ and $\actdec{z}$ which correspond to reaching consensus on value one or value zero and action $\actdec{b}$.
The action $\actdec{b}$ requires at least one process starting in location $x_0$ and one in $x_1$, and $pk = 1$, \ie an honest king round.
If $\actdec{b}$ is fired, processes can transition to an arbitrary output location.

The correctness properties of king consensus are
\begin{itemize}
    \item \textbf{King Persistency}: If all honest players have the same input value $x$, then all honest players output $y_i = x$, even if the king is faulty. We translate this to $x_1 = 0 \Rightarrow \globally(k_1 = 0)$ and $x_0 = 0 \Rightarrow \globally(k_0 = 0)$.
    \item \textbf{King Consistency}: If the king $\player_k$ is honest, then all honest players output the same value $y \in \{0,1\}$. We translate this to $pk = 0 \Rightarrow \globally(k_0 = 0 \vee k_1 = 0)$.
\end{itemize}

\subsection{Benchmark Setup}\label{sec:bench-setup}

All experiments have been conducted on a machine with two AMD EPYC 7773x processors, with 64 cores each and 2TB RAM, running Ubuntu 24.04.5 as host OS with kernel version 5.15.0.
As SMT solver Z3~\cite{z3} v.4.16.0 has been used.

The reported execution time is the measured wall clock time rounded to two decimal places.
To measure the memory consumption, our benchmarking script samples the sum of \texttt{VmRSS}, \ie the reported resident set size, of all child processes every 1ms, and takes the maximum with the reported \texttt{VmHWM}, \ie the reported peak resident set size.

\end{document}